\documentclass{aa}

\usepackage{natbib}
\usepackage{graphicx}
\usepackage{longtable,lscape}   
\usepackage{ulem}   
\usepackage{color}
\usepackage{txfonts}
\newcommand{\tess}{{\it TESS}}

\newcommand{\kms}{km\,s$^{-1}$}
\newcommand{\jktebop}{{\sf JKTEBOP}}

\definecolor{dgreen}{rgb}{0.1, 0.53, 0.22}

\begin{document}%
\title{TESS photometry and CAOS spectroscopy of six eclipsing binaries with Am components}

   \author{G. Catanzaro
          \inst{1}
          \and
          A. Frasca \inst{1}
          \and
          J. Alonso-Santiago\inst{1}
          \and
          C. Colombo\inst{2}
          }

   \institute{INAF--Osservatorio Astrofisico di Catania, via S. Sofia 78, 95123 Catania, Italy\\
              \email{giovanni.catanzaro@inaf.it}
         \and
             Universit\'a degli Studi di Catania, Dipartimento di Scienze Biologiche, Geologiche e Ambientali, Via Androne 81, I-95124 Catania, Italy\\
                         }

   \date{Received ---; accepted ---}


  \abstract{In this paper, we present the results of a comprehensive study of six eclipsing binaries whose components are confirmed or suspected Am stars. By combining long-term high-resolution CAOS spectroscopy and \tess\ photometry we have been able to accurately obtain the orbital parameters of each system as well as the atmospheric parameters of its components. We performed an in-depth chemical analysis and provided chemical abundances of C, O, Na, Mg, Si, Ca, Sc, Ti, Cr, Mn, Fe, Ni, Zn, Sr, Y, Zr, and Ba. From the solution of the light and radial curves, we have determined masses, radii, and temperatures with good accuracy. We observe apsidal motion in the eccentric system HD\,216429, in which the Rossiter-McLaughlin effect is also noted. We inferred the age of our targets by fitting isochrones on the HR diagram and find that both components in each system are properly described with the same isochrone, which reinforce our results. Furthermore, dynamical and evolutionary masses, independently obtained, show an excellent agreement.
  According to the out-of-eclipse variability shown in their \tess\ light curves and their position on the HR diagram, we claim the pulsating nature of the stars HD\,42954 (as $\delta$~Sct type) and HD\,151604 ($\gamma$~Dor).
  Based on the chemical analysis we corroborate that four of the systems studied here are formed by Am stars, while in the remaining ones (HD\,126031 and HD\,216429) only the primary component exhibits a peculiar composition. Additionally, the age distribution found in Am stars supports their suitability as age tracers in stellar populations.}
 

   \keywords{binaries: eclipsing -- binaries: spectroscopic -- stars: chemically peculiar -- stars: fundamental parameters -- stars: abundances -- stars: individual: HD\,42954, HD\,46052\, HD\,126031, HD\,151604, HD\,195020, HD\,216429}     

   \maketitle
%


\section{Introduction}
Among stars on the main sequence, A-type stars exhibit a wide array of distinct chemical characteristics. Various physical processes, like diffusion and/or magnetic fields, drive these peculiarities. They all share a common factor: a highly stable radiative atmosphere, a crucial condition for these peculiarities to manifest.

The Am stars, or metallic stars, deviate from the standard in a noteworthy way. Their \ion{Ca}{ii}~K-line types appear too early for their hydrogen line types, and their metallic-line types appear too late, resulting in a difference of five or more spectral sub-types between the inferred spectral types from Ca II K-lines and metal lines. Marginal Am stars, on the other hand, show a difference of less than five sub-types between Ca II K-lines and metal lines.
In the commonly used classification for this star class, three spectral types are prefixed with {\it k}, {\it h}, and {\it m}, corresponding to K-line, hydrogen lines, and metallic lines, respectively. The typical abundance pattern indicates lower levels of C, N, O, Ca, and Sc, and higher levels of the Fe-peak elements, Y, Ba, and rare earth elements \citep[and references therein]{catanzaro2019}.

\begin{table}
    \caption{List of the studied targets. We reported HD number, other ID, equatorial coordinates (J2000), visual apparent magnitude ($V$), and number of observed spectra (N).}
    \label{logbook}
    \centering
    \begin{tabular}{cccccc}
    \hline
    \hline
    \noalign{\smallskip}
        HD & ID & RA & DEC &$V$ & N\\
    \hline
    \noalign{\smallskip}
        \object{42954}  &   \dots   & 06 14 28.6 & +17 54 22.9 & 8.5  & 22\\ 
        \object{46052}  & WW\,Aur   & 06 32 27.2 & +32 27 17.6 & 5.8  & 17\\
        \object{126031} & DV\,Boo   & 14 22 49.7 & +14 56 20.1 & 7.5  & 16\\
        \object{151604} & V916\,Her & 16 46 35.5 & +41 47 32.2 & 7.9  & 23\\
        \object{195020} & MP\,Del   & 20 28 26.6 & +11 43 14.5 & 7.6  & 19\\
        \object{216429} & V364\,Lac & 22 52 14.8 & +38 44 44.6 & 8.4  & 19\\
    \hline
    \end{tabular}
\end{table}

Years ago \citep[see][]{1998A&A...331..627L,catanzaro2006}, we launched at the Catania Astrophysical Observatory a detailed observational campaign focusing on stars listed in the "General catalog of Ap and Am stars" \citep{1991A&AS...89..429R,2009A&A...498..961R}. The goals of this project are two-fold: confirm elemental peculiarities (if any) through abundance analyses and detect potential stellar companions by gathering radial velocity data. The effort aimed to enhance our understanding of peculiar stars and binary systems brought to several papers in the recent literature \citet{catanzaro2015,2016MNRAS.460.1999C,catanzaro2019,2020RAA....20..167F,2020MNRAS.499.3720C,2022MNRAS.515.4350C}. 
In this context, we present in this study a thorough analysis of six SB2 eclipsing binaries (listed in Table~\ref{logbook}), characterized by components belonging or suspected to belong to the Am subclass. Our analysis encompasses both high-resolution spectroscopy and photometry for a comprehensive understanding of these systems.

This paper is organized as follows. In Sect,~\ref{obs} we present our spectroscopic observations along the \tess\ archival data used. Their analysis and subsequent results are shown in Sect.~\ref{Sec:Anal}. In Sect.~\ref{sec:hr} we infer the age of our targets from their positions on the Hertzsprung-Russell (HR) diagram while their chemical composition is assessed in Sect.~\ref{sec:chem}. We discuss our results in Sect.~\ref{sec:discuss} for each individual object. Finally, we summarize the main results and the conclusions of our work in Sect.~\ref{sec:concl}.

\section{Observations and data reduction}
\label{obs}

\subsection{Spectroscopy}
\label{Subsec:Obs_spec}
Time-resolved spectroscopy of our sample of stars was carried out with the Catania Astrophysical Observatory Spectropolarimeter (CAOS) which is a fiber-fed, high-resolution, cross-dispersed echelle spectrograph \citep{2016AJ....151..116L} mounted at the Cassegrain focus of the 91 cm telescope of the ``M. G. Fracastoro'' observing station of the Catania Astrophysical Observatory (Mt Etna, Italy).

\begin{figure}
    \begin{center}
	\includegraphics[width=\columnwidth]{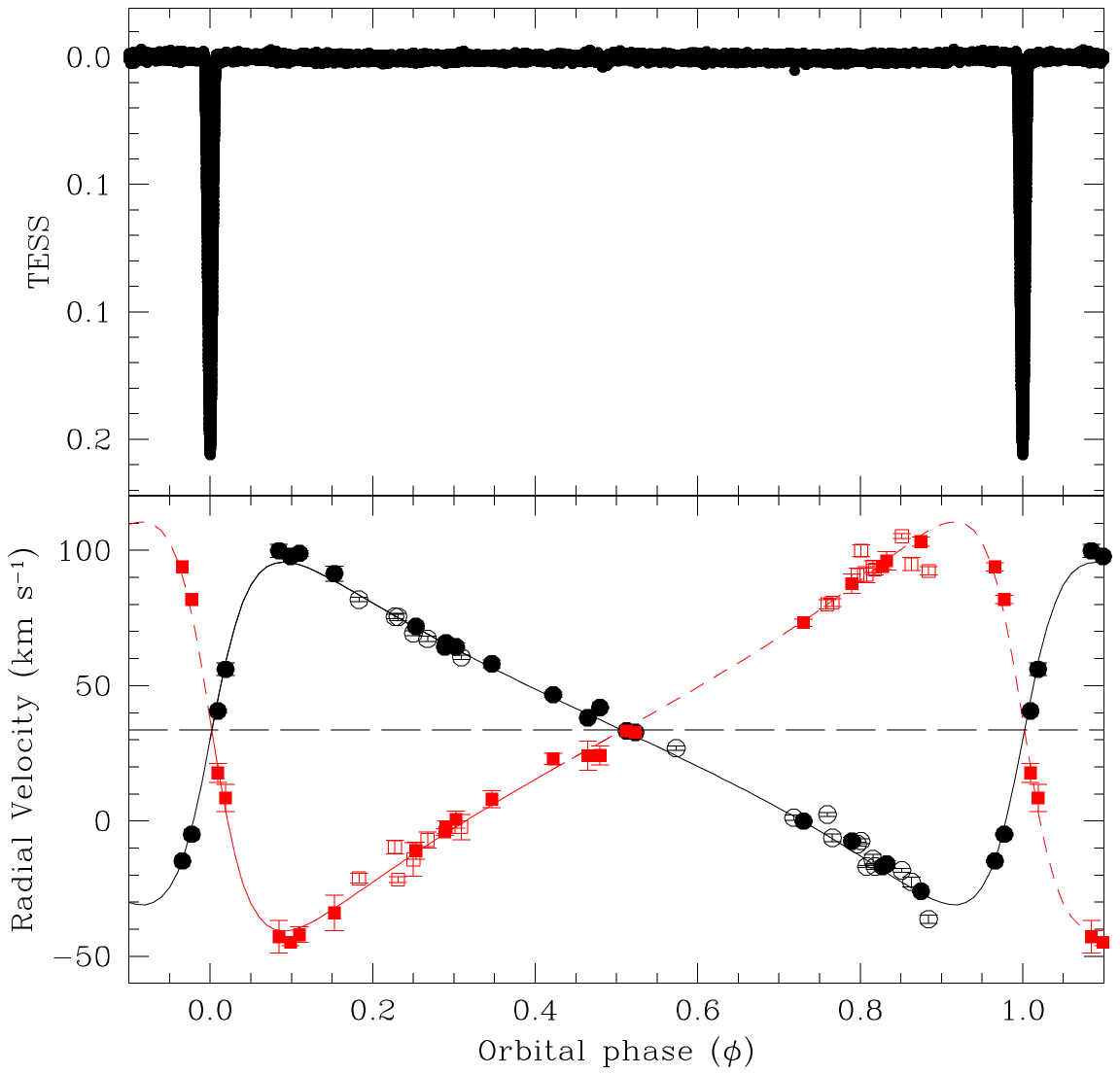}
    \caption{\tess\ light curve (top panel) and RV curve (bottom panel) for HD\,42954. The meaning of colors and symbols is the following: 
    black circles and red squared represent velocities for primary and secondary components respectively, while filled symbols represent 
    data by our work and open symbols by \citet{abt85}.}
    \label{fig:orbit_HD42954}
    \end{center}
\end{figure}

Our spectra were obtained in a large period spanning eight years, precisely from March 2014 to September 2022. Exposure times were chosen to achieve a signal-to-noise ratio (S/N) of at least 100 in the continuum in the 3900\,$-$\,6800 {\AA} spectral range. The final resolution is R\,=\,$\lambda/\Delta\lambda$ = 45\,000, as measured from ThAr and telluric lines.

The reduction of the spectra, which included the subtraction of the bias frame, trimming, correcting for the flat field and the scattered light, order extraction, and wavelength calibration, was done using the NOAO/IRAF\footnote{IRAF is distributed by the National Optical Astronomy Observatory, which is operated by the Association 
of the Universities for Research in Astronomy, Inc. (AURA), under cooperative agreement with the National Science Foundation.} packages. As the determination of effective temperatures is based on the analysis of Balmer lines, we paid 
special attention to the normalization of the corresponding spectral orders. In particular, we divided the spectral order containing H${\alpha}$ and H${\beta}$ by a pseudo-continuum obtained combining the continua of the previous and subsequent echelle orders, as already outlined by \citet{catanzaro2015}. The IRAF package {\sc rvcorrect} was used to determine the heliocentric velocity correcting the spectra for the Earth's motion.

\begin{figure}
    \begin{center}
	\includegraphics[width=\columnwidth]{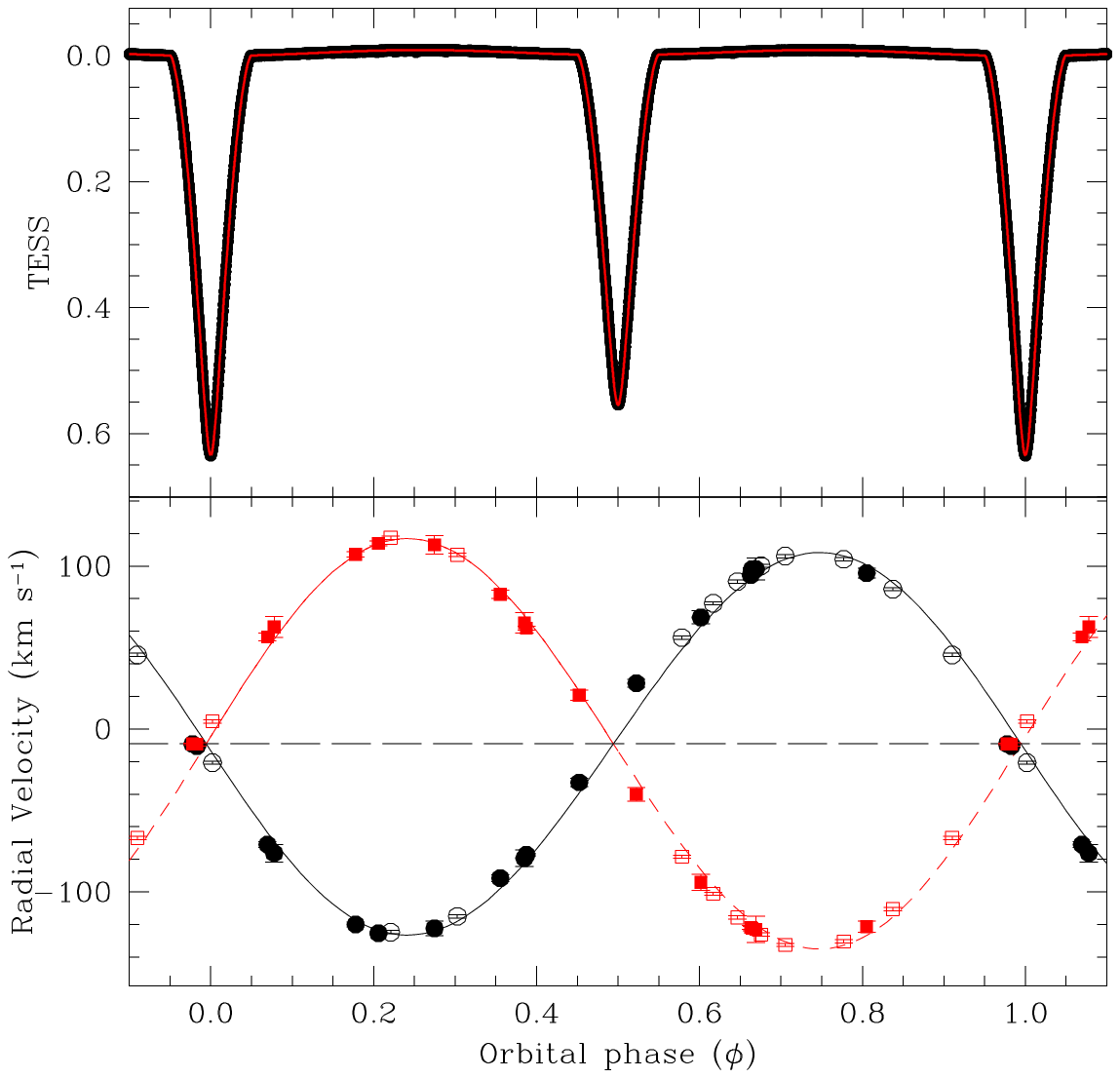}
    \caption{Light and RV curves for HD\,46052. In the top panel, the red line is the model computed with \jktebop. In the bottom panel, symbols and colors are the same as in Fig.~\ref{fig:orbit_HD42954}, but for this star literature data are from \citet{takeda19}.}
    \label{fig:orbit_HD46052}
    \end{center}
\end{figure}

\begin{figure}  
    \begin{center}
	\includegraphics[width=\columnwidth]{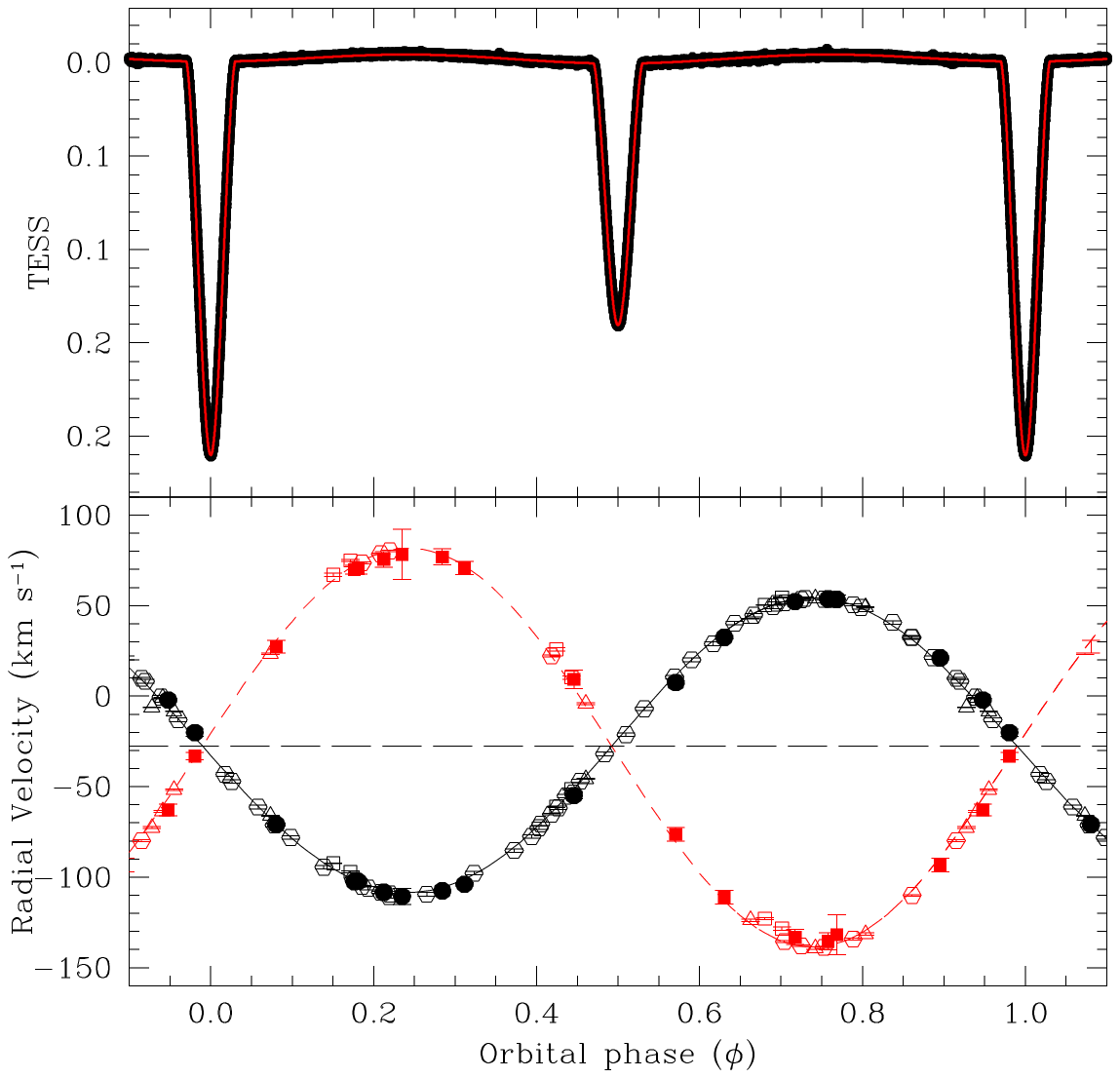}
    \caption{Light and RV curves for HD\,126031. In the top panel, the red line is the model computed with \jktebop. In the bottom panel, symbols and colors are the same as in Fig.~\ref{fig:orbit_HD42954}. Literature data are from \citet{filiz2020} (squares for FEROS and triangles for HARPS data) and from \citet{2004MNRAS.352..708C} (hexagons).}
    \label{fig:orbit_HD126031}
    \end{center}
\end{figure}

\begin{figure}
    \begin{center}
	\includegraphics[width=\columnwidth]{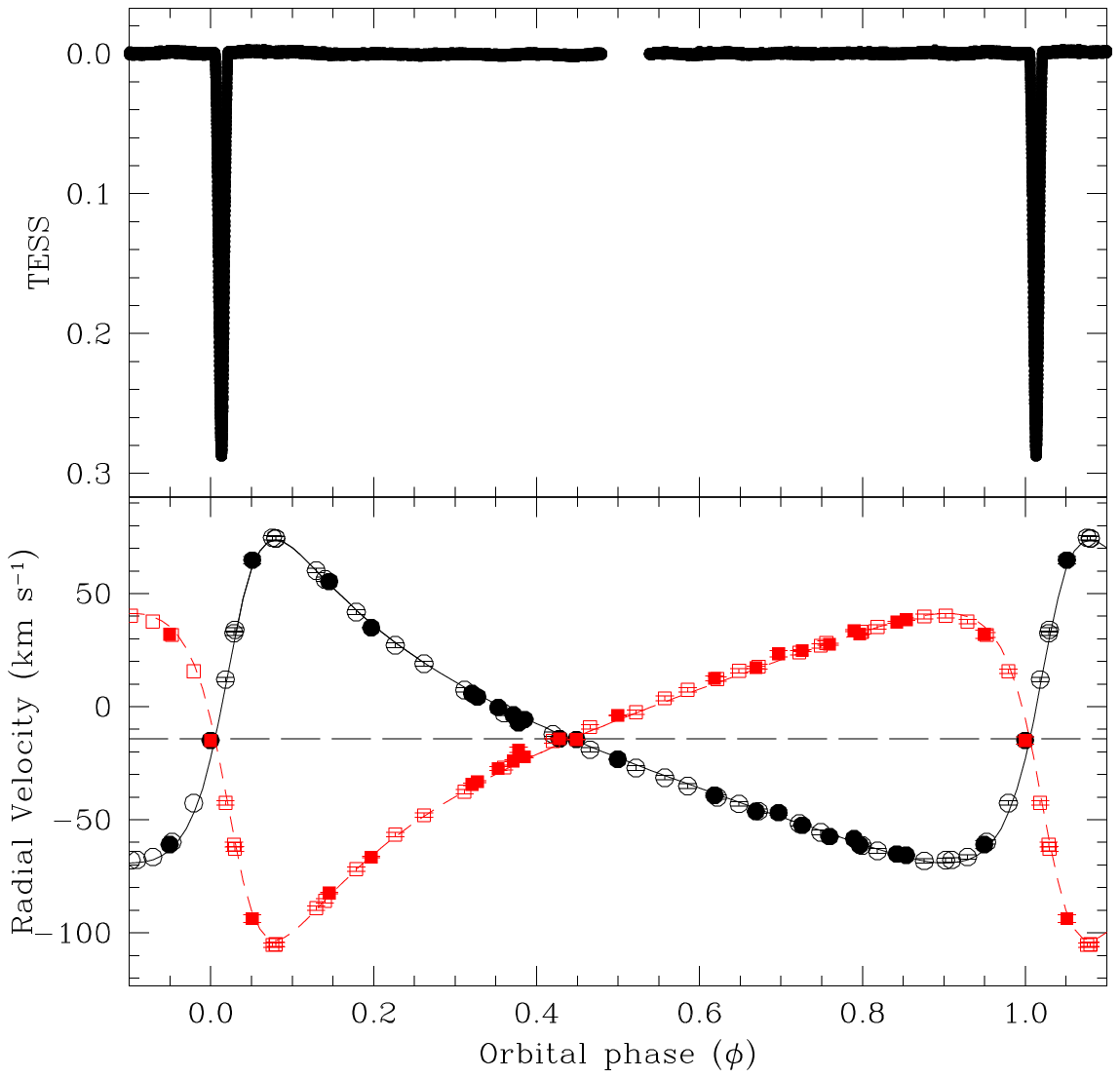}
    \caption{Light and RV curves for HD\,151604. Symbols and colors are the same as in Fig.~\ref{fig:orbit_HD42954}. Literature velocities are from \citet{2007MNRAS.380.1064C}.}
    \label{fig:orbit_HD151604}
    \end{center}
\end{figure}

For each object, radial velocities have been measured by cross-correlating the observed spectra with a synthetic template computed as described in Sect.~\ref{atmo}. The cross-correlation
has been calculated by the {\sc fxcor} IRAF package, paying particular attention to excluding Balmer lines from the correlation, as well as intervals with telluric lines.

To evaluate statistical and systematic errors in our radial velocities, during each night we observed stars with constant and well-known radial velocities. For any of these stars, we have computed the average radial velocity and standard deviation. Systematic errors are quantified with the average difference between radial velocity values from the literature and our measurements. These corrections are not significant if compared with the errors on radial velocities computed with {\sc fxcor}, so they have been neglected. The values of velocities for both components of our SB2 systems, jointly with their experimental errors and the heliocentric Julian date of the observations, have been reported in Table~\ref{rv_SB2}.

\subsection{Photometry}
\label{Subsec:Obs_photo}

Space-born accurate photometry was obtained with  NASA's Transiting Exoplanet Survey Satellite (\tess\; \citealt{Ricker2015}).
We downloaded the light curves (LCs) produced with the \tess\ Science Processing Operations Center (SPOC) pipeline \citep[][]{Caldwell2020} from the MAST\footnote{\url{https://mast.stsci.edu/portal/Mashup/Clients/Mast/Portal.html}} archive, which have a two-minute cadence. We used the simple aperture photometry (SAP) flux for the first three systems and the Pre-search Data Conditioning SAP flux (PDCSAP), where long-term trends have been removed, for the last ones. Indeed, for the latter systems, the PCDSAP light curves appear slightly cleaner than the SAP ones, while for the first three systems, the PDCSAP data seemed to show some artifacts.
Data in different {\tess\ sectors were used, whenever available, to improve the determination of the orbital period by measuring as many minima as possible. The best LCs were used for the eclipse solutions.

\section{Data analysis and results}
\label{Sec:Anal}

\subsection{Orbital parameters}
\label{Subsec:Orbital}
For all the stars of our sample, we obtained time-resolved spectroscopy to ascertain if they belong to a binary system by radial 
velocities modeling.

The radial velocities for a spectroscopic binary system are given from the following equation:

\begin{equation}
\label{radteo}
V_{\rm rad} = \gamma + K [\cos(\theta + \omega) + e \cos \omega]
\end{equation}

\noindent where $\gamma$ is the radial velocity of the center of mass,
$e$ is the eccentricity of the orbit, $\omega$ is the longitude of the periastron,
$\theta$ is the angular position of the star measured from the center of mass at
a given instant and $K$ is the semi-amplitude of the velocity curve as given by the
formula:

\begin{equation}
\label{asini}
K = \frac{2\pi a\sin i}{P_{\rm orb}\sqrt{1-e^2}}
\end{equation}

\noindent where $P_{\rm orb}$ is the orbital period of the system, $a$ is the semi-major axis and $i$ is the
inclination angle. 

By using velocities derived from our spectra (plus literature values if any), orbital elements have been determined by a weighted least-squares fitting to Eq.~\ref{radteo}. As usual, errors have been estimated as the variation in the parameters which increases the $\chi ^2$ of a unit. The starting value for $P$ was evaluated using the {\it Phase Dispersion Method} \citep{1978ApJ...224..953S} as coded in the NOAO/IRAF package. Table~\ref{tab:param} reports the determined orbital elements with their errors. The light and RV curves for the six stars studied here are displayed in Figs.~\ref{fig:orbit_HD42954} to \ref{fig:orbit_HD216429}, respectively.

\subsection{Light curve solution}   
\label{Subsec:photo}

The first parameter we measured with the \tess\ LCs is the orbital period, $P_{\rm orb}$. 
To this aim, we determined the times of primary and secondary minima (whenever observable) with the method of eclipse bisector \citep[e.g.,][]{Baroch2021,Frasca2023}, which allows us to reduce the uncertainties on the mid-eclipse epochs due to the measurement errors and data sampling. 
The errors of the mid-eclipse epochs, measured as the standard deviation of the values of the eclipse bisector, range from about 5 to 70\,s, depending on 
the S/N, the eclipse duration, and other instrumental or intrinsic variations, such as stellar pulsations, which affect the shape of the eclipses. 
The epochs of minima allow us to find the constant-period ephemeris that gives the best match to the data and normally improve the determinations based on the RVs. 

\begin{figure}
    \begin{center}
	\includegraphics[width=\columnwidth]{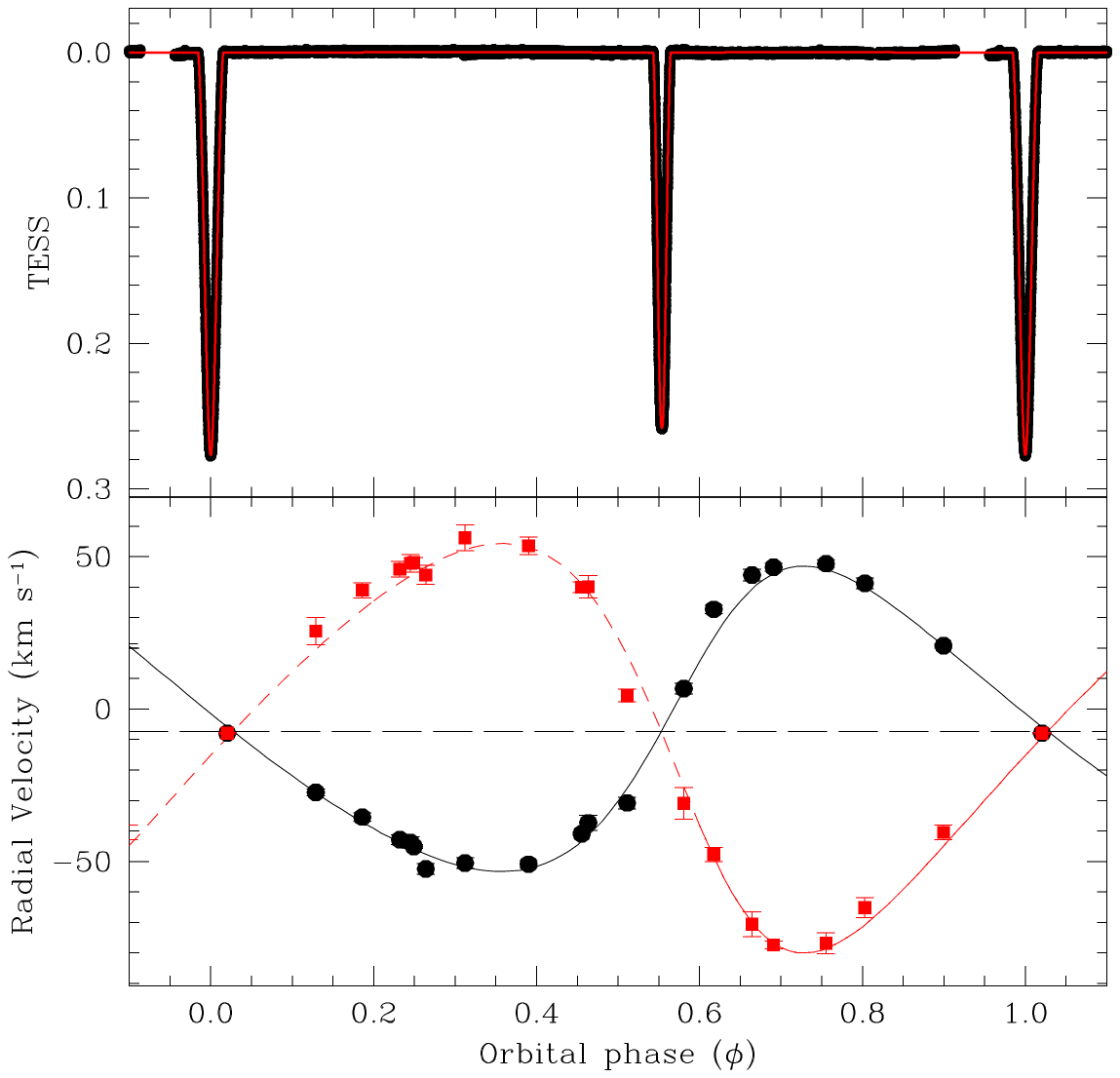}
    \caption{Light and RV curves for HD\,195020. In the top panel, the red line is the model computed with \jktebop. In the bottom panel, symbols and colors are the same as in Fig.~\ref{fig:orbit_HD42954}.}
    \label{fig:orbit_HD195020}
    \end{center}
\end{figure}

\begin{figure}
    \begin{center}
	\includegraphics[width=\columnwidth]{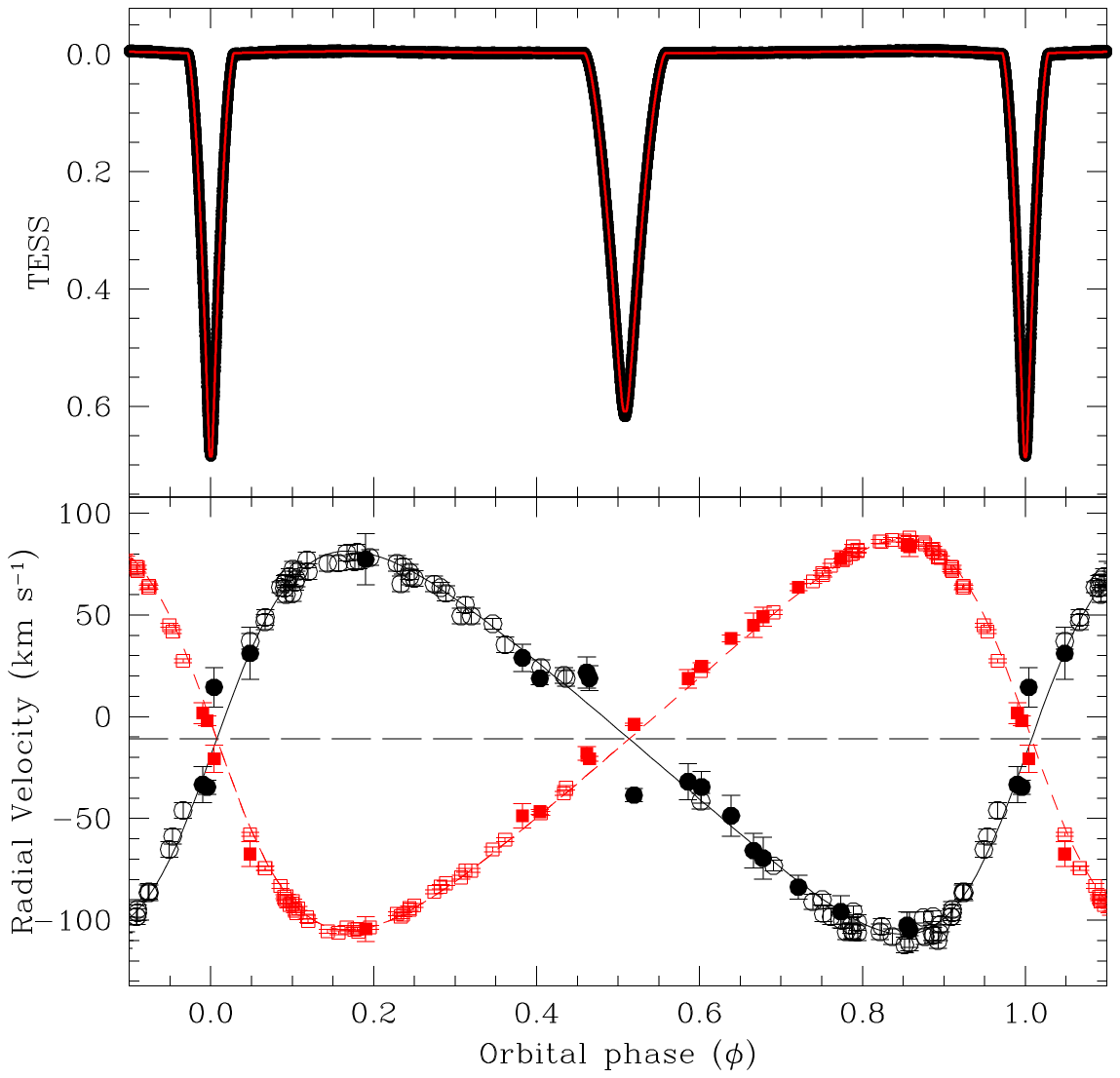}
    \caption{Light and RV curves for HD\,216429. In the top panel, the red line is the model computed with \jktebop. In the bottom panel, symbols and colors are the same as in Fig.~\ref{fig:orbit_HD42954}. Literature velocities are from \citet{torres99}. The Rossiter-McLaughlin effect is clearly visible around $\phi \approx$~0.5.}    \label{fig:orbit_HD216429}
    \end{center}
\end{figure}

As apparent in Fig.~\ref{Fig:o-c}, the residuals between observed and calculated epochs (O--C) of primary minima (blue dots) are flat and scattered around zero for all the sources, which means no relevant period variation occurred during the time spanned by the \tess\ observations.
Except for HD~216429, the O--C for secondary minima (red circles) are also flat and scattered around the time difference between the primary and secondary eclipses. The latter is zero for circular orbits because we have used $T_0+0.5\cdot P_{\rm orb}$ as the reference epoch for the secondary minima.
As pointed out, e.g., by \citet{Baroch2021}, a different slope for the O--C of the primary and secondary minima for systems with eccentric orbits is a sign of apsidal motion. This effect is visible in the O--C of HD~216429, for which the difference between secondary minima ($T_0^S$) and primary minima ($T_0^P$) changes from $0.5\cdot P_{\rm orb}$+0.063 days to $0.5\cdot P_{\rm orb}$+0.056 days between the two epochs of \tess\ observations.
HD~195020 was observed only in two consecutive \tess\ sectors and the time span of about 43 days is too small to show a clear trend. 
The other two systems with eccentric orbits, HD~42954 and HD~151604, display only one set of minima (those closest to the periastron passage), with no detection of secondary eclipses.
For the two systems with circular orbits the O--C of the secondary minima display a tiny offset for the primary ones, which could indicate a very small eccentricity that is not detectable from the RV curves. However, these differences are always within the 1$\sigma$ error bars and therefore are not significant. 

For the systems that show both primary and secondary minima, we modeled their \tess\ LCs, by using the {\sc fortran} code \jktebop\footnote{\url{https://www.astro.keele.ac.uk/jkt/codes/jktebop.html}} 
\citep{Southworth2004MNRAS,Southworth2013}. This code uses a bi-axial ellipsoidal model \citep{NelsonDavis1972}  for calculating proximity effects and the Levenberg-Marquardt optimization algorithm to find the best-fitting model, which makes it particularly suitable for fast modeling of detached eclipsing binaries like those analyzed in the present work. We ran \jktebop\ on the best LCs of our binaries, merging different \tess\ sectors for the systems with the longest periods. We used a quadratic limb-darkening law with the coefficient derived by \citet{Claret2017} for the \tess\ pass-band. The fractional stellar radii (in units of the separation), the system inclination, and the surface brightness ratio $J$ were left as free parameters in the fit. The 1$\sigma$ uncertainties in the fitted parameters were determined through 1000 Monte Carlo simulations. 
With the stellar separation derived from the solution of the RV curves, we obtained the stellar radii in absolute units. 
For these systems, we derived the temperature of the cooler component from that of the hotter, which was measured with the spectra synthesis (see Sect.\,\ref{atmo}) and the surface flux ratio $J$ measured by \jktebop. For this task, we calculated the flux ratios by integrating BT-Settl models \citep{Allard2012} through the \tess\ pass-band.

For those with only one eclipse, namely HD\,42954 and HD\,151604, we report the value of $T_{\textrm{eff}}^{\textrm{B}}$ derived with the spectral synthesis. In these cases, as the light curve could not be solved, no system inclination nor stellar radii could be derived.
The orbital and stellar parameters derived with \jktebop\ and the solution of the RV curves are listed in Table~\ref{tab:param}.
For the two systems without LC solution, we report in Table~\ref{tab:param} the dynamical masses and semi-major axis ($M_{\textrm{A,B}}\sin^3i$ and $a\,\sin\,i$) that can be obtained from the RV solution only. We note that, as these binaries exhibit at least one eclipse and their separations are much larger than the stellar radii, the orbital inclinations should be close to 90\degr, making the $\sin^3i$ factors close to unity. Moreover, for these systems, the stellar luminosities were estimated from the combined $V$ magnitude, the luminosity ratio of the components ($l$), and the $Gaia$-DR3 parallax \citep{GaiaDR3} as in \citet{Frasca2022}. The stellar radii reported in Table~\ref{tab:param} are derived from luminosities and effective temperatures.

\subsection{Atmospheric parameters}
\label{atmo}
\begin{figure*}
\centering
	\includegraphics[width=15cm,bb=0 0 504 300]{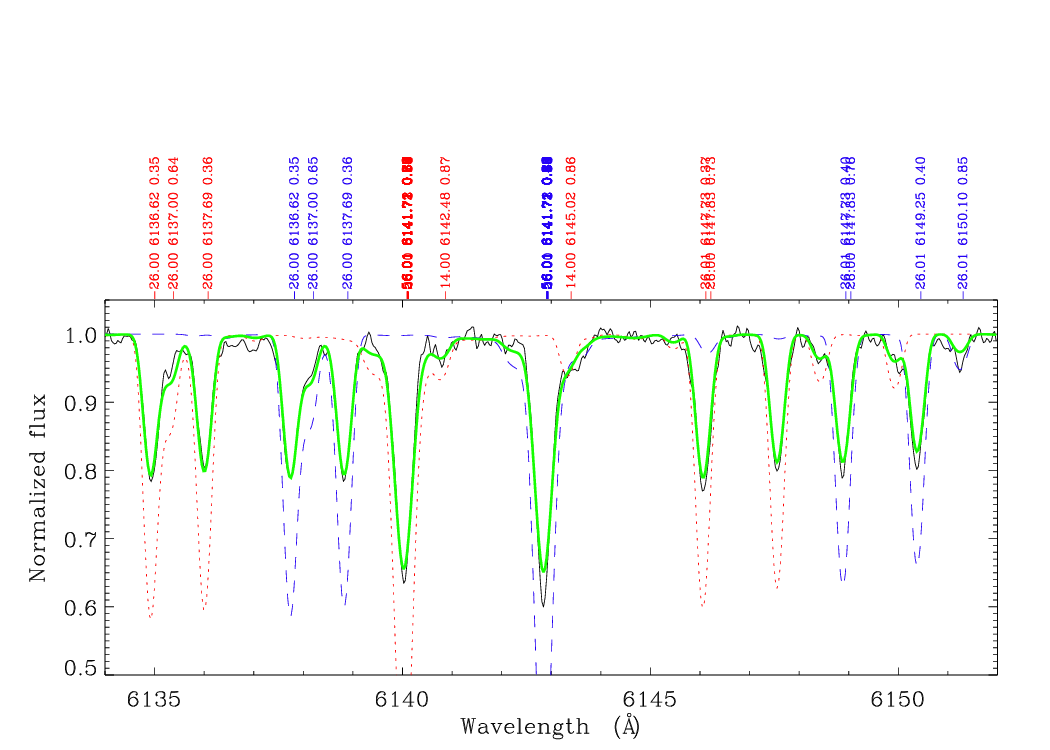}
    \caption{Results of the fitting procedure for HD\,151604 in the spectral range between $\lambda \lambda$ 6134--6152 {\AA}. The synthetic lines for the primary component are represented by blue dashed lines, while the synthetic lines for the secondary component are represented by red dotted lines. The solid green line depicts the synthetic composite spectra.}
    \label{example_figure}
\end{figure*}

One commonly used method for determining the effective temperature ($T_{\rm eff}$) of a star is to compare the observed and theoretical profiles of one or more Balmer lines. In the case of our SB2 systems, we observed H$\alpha$ and H$\beta$ profiles that were obtained by superimposing lines Doppler-shifted by the orbital motion and weighted by the different luminosities.

The procedure we employed for all our stars involved minimizing the difference between the observed and synthetic H$\alpha$ and H$\beta$ profiles obtained at the quadrature orbital phases. Following \citet{2015A&A...581A.129T} we consider all possible combinations of synthetic primary and secondary spectra derived from the computed grid. These combinations were utilized to construct composite theoretical spectra using the following formula:

\begin{equation}
\label{flux}
F^{th}_{\text{Tot}} = \frac{l_\lambda F^{th}_A + F^{th}_B}{1+l_\lambda}
\end{equation}

\noindent
where the dilution factor {\it l$_\lambda$} is defined as:

\begin{equation}
\label{dilution}
l_\lambda = \frac{I^{cont}_{\lambda,A}}{I^{cont}_{\lambda,B}} \left( \frac{R_{\textrm{A}}}{R_{\textrm{B}}} \right) ^2 
\end{equation}

\noindent
$F^{th}_A$ and $F^{th}_B$ denote the synthetic spectra (normalized to unity), I$^{cont}_{\lambda,A}$ and I$^{cont}_{\lambda,B}$ are the specific intensities at a given wavelength in the continuum and $R_{\textrm{A}}$ and $R_{\textrm{B}}$ are the radii, respectively for A and B components. In our study, $l_\lambda$ was directly computed by using derived radii from the LCs and the ratio between specific intensities in the continuum estimated from synthetic spectra. However, for HD\,42954 and HD\,151604, $l_\lambda$ was treated as a free parameter in the fitting procedure.

The fitting of H$\alpha$ and H$\beta$ profiles was carried out in three steps: first, we computed the stellar atmosphere model for solar chemical abundances using the ATLAS9 code \citep{1993ASPC...44...87K}; then, the stellar spectrum was synthesized using SYNTHE \citep{1981SAOSR.391.....K}; finally, instrumental and rotational broadening effects were applied. The rotational velocities (v\,$\sin\,i$) of our targets by matching metal lines to synthetic profiles in our highest S/N spectra were taken at the quadrature phase just to ensure the maximum separation among lines of both components. The best fit occurred for the values reported in Table~\ref{tab:param}.
Initially, models were computed using the solar opacity distribution function (ODF) table. Regarding the determination of gravities (log\,$g$), since Balmer line profiles are not sensitive to gravities, we used the values derived from the LC solution, except for HD\,42954 and HD\,151604 for which we utilized the ionization equilibrium method between \ion{Fe}{I} and \ion{Fe}{II}. Microturbulent velocities ($\xi$) were derived by using the empirical calibration as a function of the stellar effective temperature given by \citet{2014psce.conf..193G}.

The simultaneous fitting of these two lines resulted in a final solution obtained as the intersection of the two $\chi^2$ isosurfaces. An important source of uncertainties arose from the difficulties in continuum normalization as it is always challenging for Balmer lines in echelle spectra. We quantified the error introduced by the normalization to be at least 100 K, which we summed in quadrature with the errors obtained by the fitting procedure.

In addition, for the four stars for which we determined the temperature of the cool companion from the \tess\ flux ratio (see previous Sect.\,\ref{Subsec:photo}), we found by spectral synthesis consistent values. Therefore, as a matter of uniformity, we have adopted the spectroscopic temperature values for all objects in our chemical analysis.

With these parameters, we estimated the metallicities for each target, and when we found abundances inconsistent with the solar ones, we repeated the procedure with the appropriate ODF.
As an example, we show in Fig.~\ref{example_figure} the results of the procedure for the star HD\,151604, while for all the other objects we refer the reader to Fig.~\ref{all_balmer}.
The final atmospheric parameters obtained for the six systems under study, with their errors, are reported in Table~\ref{tab:param}.

\section{HR diagram}
\label{sec:hr}

\begin{figure*}
    \begin{center}
	\includegraphics[width=16cm]{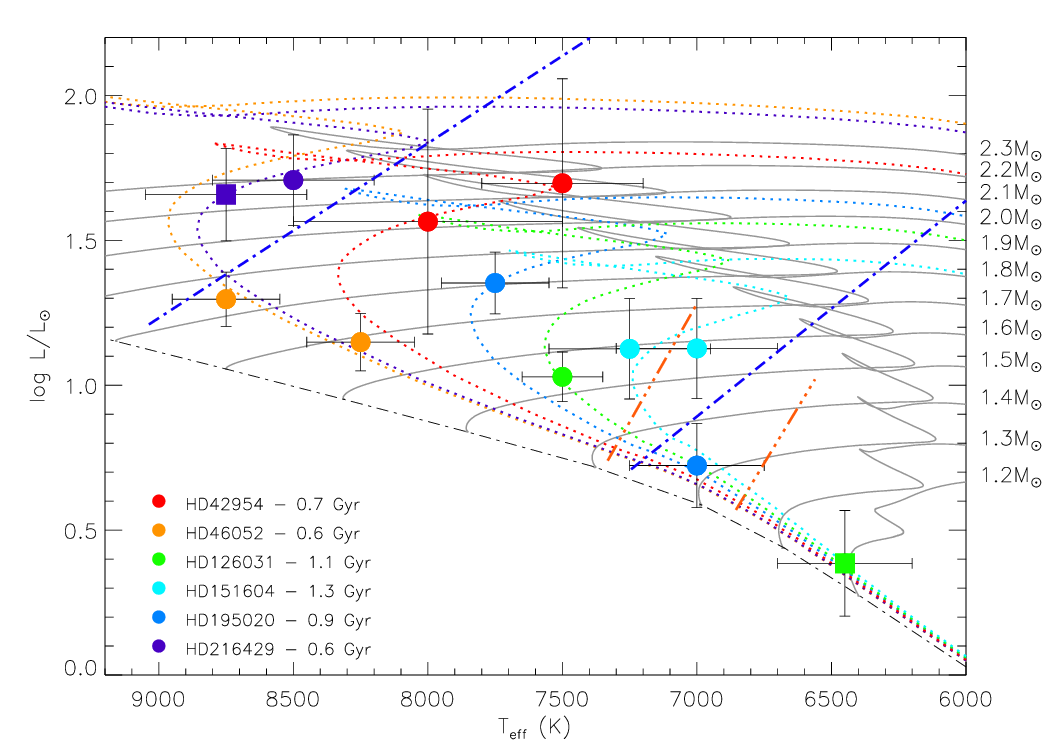}
    \caption{HR diagram for the six SB2 systems investigated in this paper, colors are different according to the object. The evolutionary tracks (solid lines) for the labeled masses as well as the ZAMS (dash-dotted line), and the isochrones (dotted lines plotted with the same color of the reference system) for solar composition. The blue dot–dashed lines show the $\delta$\,Sct instability strip by \citet{1998A&A...332..958B}; the orange dash-dot-dot-dot lines show the theoretical edges of the $\gamma$\,Dor instability strip by \citet{2003ApJ...593.1049W}. The two stars plotted as filled squares have standard solar composition.}
    \label{hr}
    \end{center}
\end{figure*}
\begin{figure}
    \centering	
    \includegraphics[width=\columnwidth]{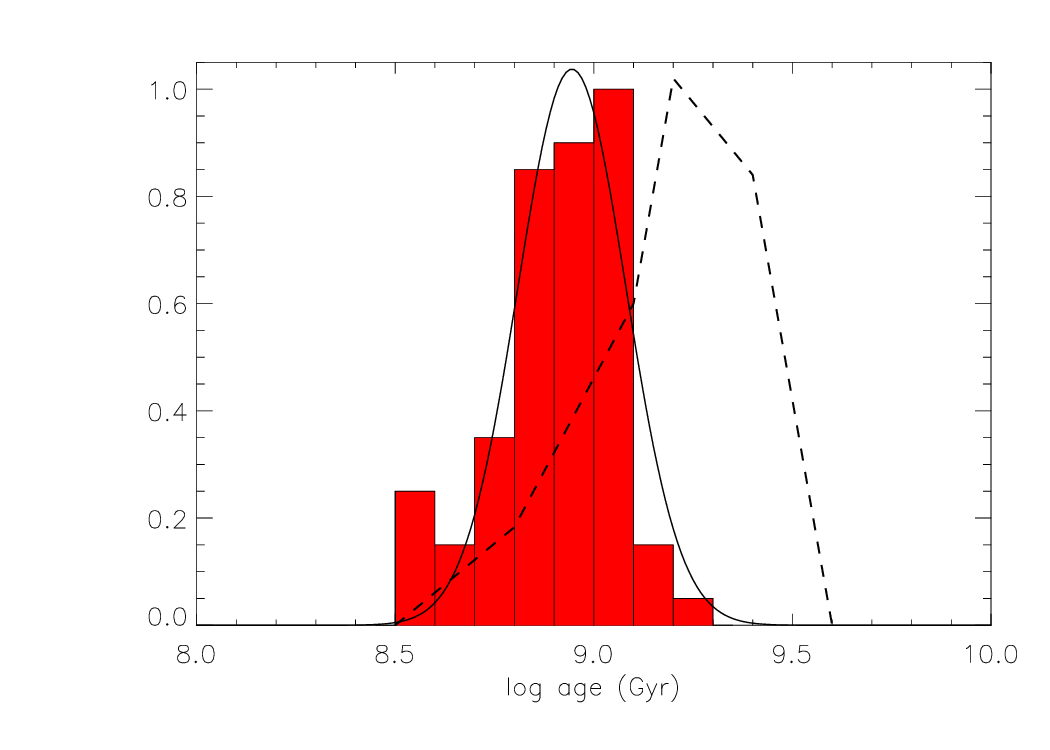}
    \caption{Comparison between the age distribution of Am stars derived in this paper (red histograms-solid line) and the one derived by \citet{worthey2015} (dashed line).}
    \label{age}
\end{figure}

\begin{figure*}
\centering
\hspace{-2.5cm}
    \includegraphics[width=14.5cm,bb=0 15 504 400]{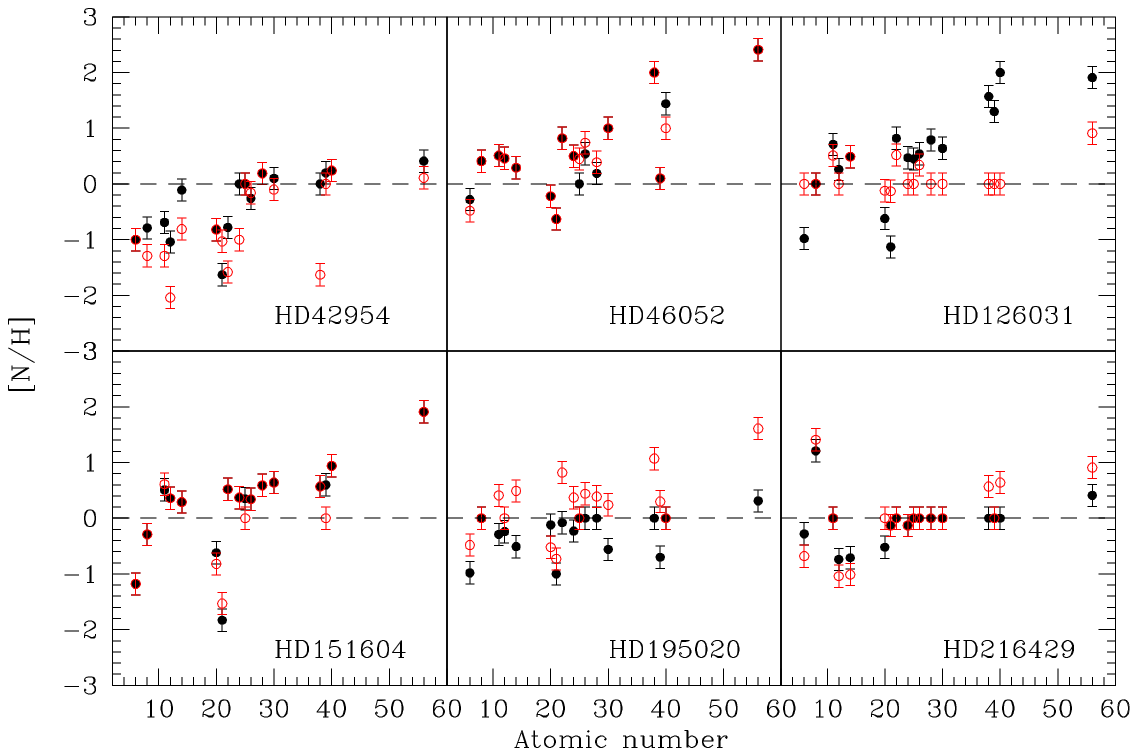}
    \caption{Chemical pattern derived for both components of our six SB2 systems. With filled black and red open circles we indicate the abundances of the primary and secondary components, respectively. The dashed line indicates the solar value.}
    \label{all_ab}
\end{figure*}

In a previous paper, \citet{2022MNRAS.515.4350C} studied a sample of five Am stars in SB1 systems to determine their ages and test the \citet{worthey2015} hypothesis of estimating the ages 
of stellar populations in galaxies using chemically peculiar (CP) objects since their contribution to the integrated spectral light of the entire population varies over time. The method is based on detecting CP star features in the light of stellar populations, as long as the S/N is sufficiently high.

We used the HR diagram (see Fig.~\ref{hr}) in order to infer the age of our targets. Luminosities ($L$) have been computed from the temperatures and radii reported in Table~\ref{tab:param} as $L=4\pi R^2\sigma T_{\textrm{eff}}^4$. We note that the largest luminosity errors are found for the two systems for which the LC could not be solved and the luminosity of their components had to be estimated as described at the end of Sect.~\ref{Subsec:photo}.

We also plotted the evolutionary tracks by \citet{2012MNRAS.427..127B} and non-rotating PARSEC v2.0 isochrones \citep{2022A&A...665A.126N} computed for solar metallicity (Z$_{\odot}$=0.0152).
All the stars are in the hydrogen-burning phase, covering an age range between $\approx$\,0.50~Gyr and $\approx$\,1.29~Gyr (see Table~\ref{tab:param}) and a mass interval between $\approx$\,1.2~M$_\odot$ and $\approx$\,2.3~M$_\odot$. It is important to note that the masses indicated for our targets from the evolutionary tracks are perfectly consistent (within the experimental errors) with the dynamical masses determined by analysis of \tess\ light curves and CAOS radial velocities. Furthermore, for all six binaries, both components are fitted by a single isochrone, reflecting the robustness of our analysis.

Additionally, the instability strips both for $\delta$~Sct \citep{1998A&A...332..958B} and for $\gamma$\,Dor \citep{2003ApJ...593.1049W} are also shown in the HR diagram to clarify the out-of-eclipse light variability of two of our targets (which will be discussed in detail in Sects.~\ref{hd42954} and \ref{hd151604} for HD\,42954 and HD\,151604, respectively).

To increase the sample of Am stars with known ages, we combined our results with those in \citet{catanzaro2019,2022MNRAS.515.4350C}, after recalculating, for consistency, the ages with the same isochrones used in this study. Results are shown in Fig.~\ref{age}, where we compare the distribution of ages derived here with that from \citet{worthey2015}. Our distribution of $\log (age)$ is centered around 8.94 with a FWHM\,=\,0.32. Thus, we confirmed that stars that can contribute to the integrated light are confined in a period between 300 million years and 3.2 billion years, as no Am-class stars with ages outside these limits are observed. However, it is evident that the two distributions have the same shape, but our data are shifted toward younger ages of about 0.35 Gyr.

\section{Chemical analysis}\label{sec:chem}
Once the atmospheric parameters have been obtained, we proceed to determine the abundances of individual species. For this we applied the spectral synthesis technique \citep{2011MNRAS.411.1167C,catanzaro2013}, as done in previous works \citep[see, e.g.,][and references therein]{2023arXiv231208581A}. We compared the observed spectra with the synthetic ones computed in Sect.\ref{atmo}, calculated using Eq.~\ref{flux}. Then, the abundances were derived by minimizing their difference. It was evaluated in 25 {\AA} wide intervals by performing a $\chi^2$ minimization implemented in IDL\footnote{IDL (Interactive Data Language) is a registered trademark of NV5 Geospatial.} using the AMOEBA routine. We utilized spectral line lists and atomic parameters from \citet{2004A&A...425..263C}.

In this way we investigated 17 elements, namely: C, O, Na, Mg, Si, Ca, Sc, Ti, Cr, Mn, Fe, Ni, Zn, Sr, Y, Zr, and Ba. All the abundances inferred for both components of our binaries, relative to the solar composition described in \citet{2010Ap&SS.328..179G}, have been reported in Table~\ref{tab:abun} and plotted in Fig~\ref{all_ab}. To reliably evaluate the uncertainty on individual abundances, we added in quadrature the standard deviation of the abundances obtained for each spectral chunk, and the uncertainty derived from the propagation of errors on the atmospheric parameters.
We noticed that the latter contribution is the dominating source of uncertainty, mainly due to the dilution factor $\delta$\,l$_\lambda$, and the overall abundance uncertainties turned out to be of the order of 0.2~dex, which, as expected, is larger than typical values for single-lined objects. 

As will be discussed in detail in Sect.~\ref{sec:discuss}, we have found Am-type peculiarities in all the components of our six SB2 systems, except for two of them for which the abundances turn out to be solar-type.

\section{Notes on individual objects}\label{sec:discuss}
In this section, we describe the results obtained for each star from both chemical and orbital points of view, comparing them with literature results, when present.

\subsection{HD\,42954 (= HR\,2214)}
\label{hd42954}
HD\,42954, initially designated as a spectroscopic binary (SB2) system by \citet{abt85}, presented challenges in accurately determining the secondary star's physical parameters due to limitations in the precision of their radial velocity measurements. Additionally, \cite{cowley69} classification identifying HD\,42954 as an Am-type star underscored intriguing chemical peculiarities within the system. However, a recent extensive analysis by \citet{2008MNRAS.389..925T} redefined HD\,42954 as a triple system, offering crucial insights into mass distribution. The primary and secondary stars were found to have masses of 2.10 and 1.67 M$_\odot$, respectively, while the tertiary star exhibited a mass of approximately 2.09 M$_\odot$. \citet{2008MNRAS.389..925T} also provided precise orbital period data: $\log P_S$\,=\,1.38~d for the inner period and $\log P_L$\,=\,4.94~d for the outer period. 
In our data, we do not find evidence of the third component of the system as hypothesized by the last author, but we have a very good agreement about both the masses of the components of the inner pair and their orbital period, albeit slightly shorter than that found by \citet{2008MNRAS.389..925T}, as reported in Table~\ref{tab:param}.

In agreement with this new understanding, our recent investigation strengthens the Am-star classification for HD\,42954. We observed a similar pattern in the distribution of abundances of both components, with general normality in the peak of iron and heavy elements, and a pronounced underabundance of $\alpha$ and light elements, particularly calcium, scandium, and titanium. We note in the secondary component abundances a discrepancy from the primary in magnesium, chromium, and strontium.

The \tess\ light curves of this system display clear out-of-eclipse variations reminiscent of pulsations. We have therefore carried out a periodogram analysis \citep{Scargle1982}, masking the eclipses, and applying the CLEAN deconvolution algorithm \citep{Roberts1987} to reduce aliases caused by the gaps in the data. We found several modes, which come out of all the \tess\ sectors analyzed independently. The cleaned periodogram corresponding to the data obtained in the two consecutive sectors 44 and 45 (12 October -- 2 December 2021) is shown in Fig.~\ref{fig:freq_HD42954}, where all the peaks higher than 1.5 times the 99\% confidence level are indicated with $f_{i}$ and their corresponding values are reported in Table~\ref{Tab:freq_HD42954}. The peaks just above the confidence level are to be considered only as possible pulsation frequencies that we have not taken into consideration to remain conservative. 

The periodogram shows an excess of power at frequencies greater than 5~d$^{-1}$. Therefore, according to the pulsation classification scheme proposed by \citet{2010ApJ...713L.192G}, HD\,42954 can be classified as a $\delta$~Sct variable. This classification is further supported by the object's position on the HR diagram (see Fig.~\ref{hr}), which is located within the instability strip characteristic of $\delta$\,Sct stars \citep{1998A&A...332..958B}.
\begin{figure}[]
\begin{center}
\hspace{-.5cm}
\includegraphics[width=\columnwidth]{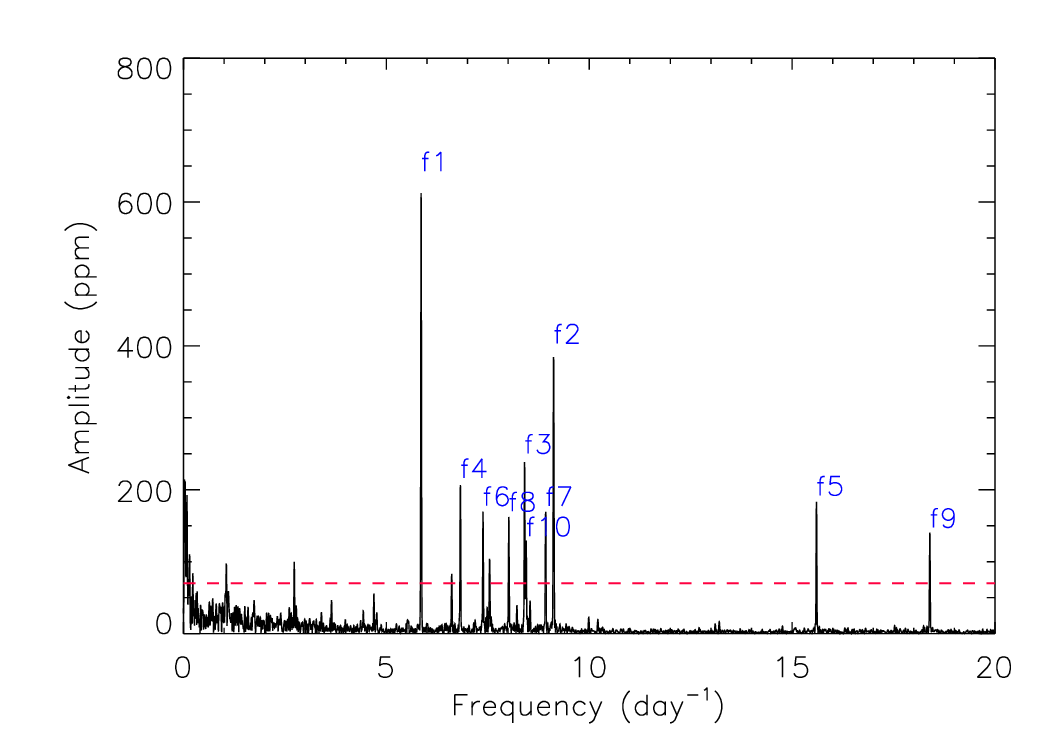}	
\vspace{0cm}
\caption{Cleaned periodogram of the \tess\ light curves of HD\,42954 taken in Sectors 44 and 45 (Oct--Nov 2021), where the eclipses have been removed. The frequency peaks overcoming the 99\% confidence level (red dashed line) by a factor of 1.5 have been indicated with $f_1$ to $f_{10}$, in order of decreasing amplitude, and their values are reported in Table~\ref{Tab:freq_HD42954}. }
\label{fig:freq_HD42954}
\end{center}
\end{figure}

\begin{table}
    \caption{Pulsation frequencies of HD\,42954.}
    \label{Tab:freq_HD42954}
    \centering
    \begin{tabular}{crc}
    \hline
    \hline
    \noalign{\smallskip}
    Mode &     Frequency  & Amplitude   \\
         & (day$^{-1}$)~   &  (ppm)   \\
    \hline
    \noalign{\smallskip}
$f_{1}$  &    5.85563  &   612 \\
$f_{2}$  &    9.12154  &   384 \\
$f_{3}$  &    8.40570  &   238 \\
$f_{4}$  &    6.82331  &   206 \\
$f_{5}$  &   15.60181  &   183 \\
$f_{6}$  &    7.38250  &   169 \\
$f_{7}$  &    8.92523  &   169 \\
$f_{8}$  &    8.01704  &   162 \\
$f_{9}$  &   18.39577  &   140 \\
$f_{10}$ &    8.44536  &   129 \\
    \noalign{\smallskip}
    \hline
    \end{tabular}
\end{table}

\subsection{HD\,46052 (= HR\,2372)}
In a recent investigation, \citet{2021A&A...652A.120I} curated a uniformly compiled dataset of main-sequence OBA-type dwarfs in eclipsing binary systems using \tess\ photometry. Among their findings, HD\,46052 was identified as an eclipsing binary system, categorizing the components as Am stars and indicating chemical peculiarities in their composition. Notably, a prior study by \citet{2019AJ....157..211M} had previously examined HD\,46052, they derived primary component parameters of $M_{\textrm{A}}$\,=\,2.297 M$_{\odot}$ and $R_{\textrm{A}}$\,=\,2.731 R$_{\odot}$, while also identifying a late spectral type for the secondary component. Another important study on this binary system was conducted by \citet{takeda19}, who characterized the two components both dynamically and chemically. Among other results, they find $M_{\textrm{A}}$\,=\,1.964 M$_{\odot}$, $M_{\textrm{B}}$\,=\,1.814 M$_{\odot}$, $R_{\textrm{A}}$\,=\,1.927 R$_{\odot}$, $R_{\textrm{B}}$\,=\,1.841 R$_{\odot}$ and, with regards to the temperatures, $T_{\textrm{A}}$\,=\,7960~K and $T_{\textrm{B}}$\,=\,7670~K.

From the analysis of the \tess\ LCs we derived also masses and radii for the components, and in particular results are in agreement with those authors. We obtained a primary component with $M_{\textrm{A}}$\,=\,1.95\,$\pm$\,0.01 M$_{\odot}$ and $R_{\textrm{A}}$\,=\,1.94\,$\pm$\,0.02 R$_{\odot}$ and a secondary component with $M_{\textrm{B}}$\,=\,1.82\,$\pm$\,0.01 M$_{\odot}$ and $R_{\textrm{B}}$\,=\,1.85\,$\pm$\,0.02 R$_{\odot}$. However, our analysis deviates from the result of \citet{2019AJ....157..211M} regarding the spectral type of the secondary component. Our findings suggest that the spectral type of the secondary component is similar to that of the primary (although, given the 500 K difference between the two temperatures, there could be at least two subclasses of difference between the two stars), calling into question the late spectral type previously proposed. Moreover, these results are not in agreement with those of \citet{takeda19}, although they confirm the same type of peculiarities. The reason for this discrepancy could be due to the different $T_{\textrm {eff}}$ of the two components found in our study, which are hotter and not consistent with those of \citet{takeda19}. However, our analysis reveals a consistency between the dynamic masses and those derived from the evolutionary tracks (see Sect.~\ref{sec:hr}), which is only possible using the temperatures we found. The same agreement is not possible with colder temperatures on the order of those obtained from the previous authors' study.

We also investigated the possible presence of pulsations in the out-of-eclipse \tess\ light curve. The periodogram depicted in Fig.~\ref{fig:freq_HD46052_HD126031} (left upper panel) shows uniquely two peaks corresponding to $f_1$\,=\,0.3968~d$^{-1}$ ($P$\,$\approx$\,2.52~d, i.e. nearly equal to the orbital period) and $f_2$\,=\,0.7934~d$^{-1}$ (half orbital period $P$\,$\approx$\,1.26~d), which are related to proximity effects (reflection and ellipsoidal shape of the components). No other power excess is present above the 99$\%$ confidence level, so we exclude that HD\,46052 may be a pulsating variable.

Our recent in-depth study corroborates previous evidence, confirming that both components indeed align with the Am stellar classification and show consistent chemical abundances among them. In particular, there are under-abundances of carbon, calcium, and scandium, and small over-abundances of light elements such as oxygen, sodium, magnesium, and silicon, while the iron peak and heavy elements generally (especially Sr and Ba) show over-abundances compared to the sun. Exceptions are manganese in component A and yttrium in both stars, which show solar abundance.

\subsection{HD\,126031 (= DV\,Boo)}
The binary system, initially identified as an Algol-type eclipsing binary by the {\it Hipparcos} mission, garnered attention in subsequent studies. \citet{1988PASP..100.1084B} classified it as an Am star, \citet{1999A&AS..137..451G} further refined the classification to A3(k) A7(h) F5(m).

\citet{filiz2020} conducted an extensive investigation, achieving remarkable precision in determining the masses of the binary components, accurate to within 1$\%$. The primary component was found to have a mass of 1.593\,$\pm$\,0.002 M$_\odot$, while the secondary component exhibited a mass of 1.188\,$\pm$\,0.002 M$_\odot$. Additionally, both stars were determined to have solar radii, ascertained with an accuracy of less than 3$\%$. The system's estimated age was reported to be 1.00\,$\pm$\,0.08 Gyr. An abundance analysis primarily focused on the primary component due to the lower S/N of the secondary component, revealing a typical Am star pattern. Additionally, the primary star was identified as a candidate $\delta$~Scuti-type pulsating star using data from All Sky Automated Survey (ASAS) archive \citep{2002AcA....52..397P}.

\begin{figure}[]
\begin{center}
\hspace{-.5cm}
\includegraphics[width=\columnwidth]{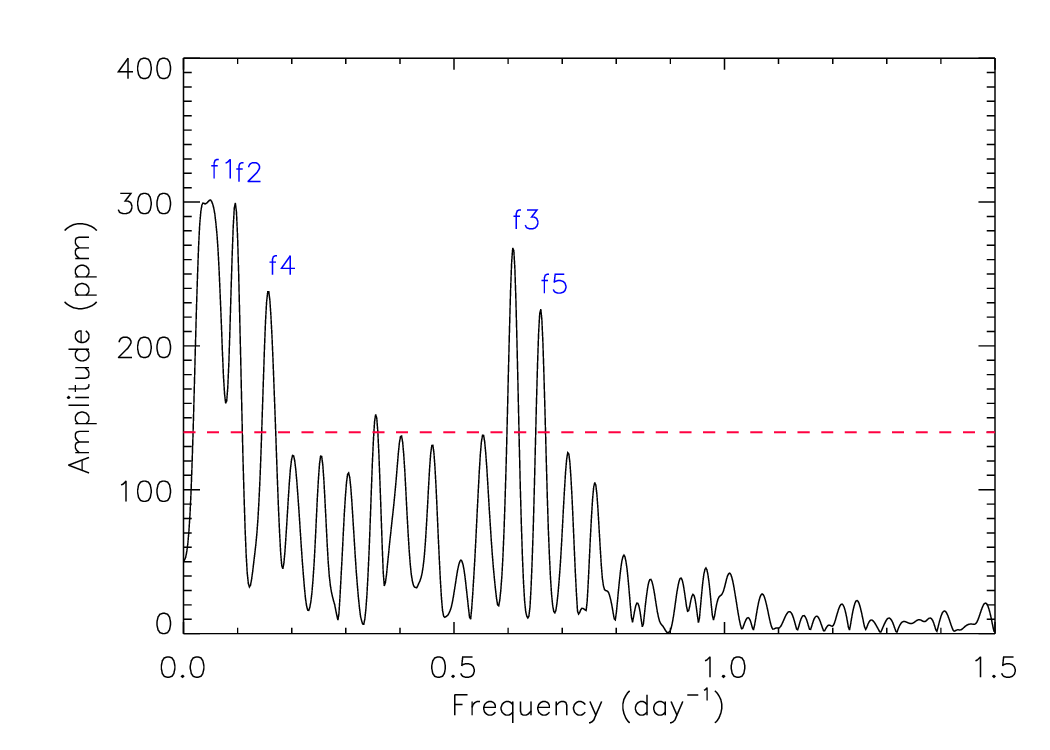}	
\vspace{0cm}
\caption{Cleaned periodogram of the \tess\ light curves of HD\,151604 taken in Sectors 51 and 52 (April--June 2022), where the eclipses have been removed. The frequency peaks higher than {\bf 1.5 times the} 99\% confidence level (red dashed line) have been indicated with $f_1$ to $f_{5}$, in order of decreasing amplitude and their values are reported in Table~\ref{Tab:freq_HD151604}.}
\label{fig:freq_HD151604}
\end{center}
\end{figure}

In our current analysis, we have derived masses and age of the components that are compatible with \citet{filiz2020} results, while our radii are larger than the solar value ($R_{\textrm{A}}$\,=\,1.94 R$_{\odot}$, $R_{\textrm{B}}$\,=\,1.25 R$_{\odot}$). Moreover, we validated and reinforced the findings of \citet{filiz2020} concerning the primary component, it is definitively an Am star with underabundances of carbon, calcium, and scandium, being the other elements normal or overabundant (i.e. Zr and Ba). Moreover, we extend the abundance analysis to the secondary component, for what we can affirm is a normal star since we found an almost normal (within the errors) abundance of elements.

We also investigated the possible presence of $\delta$\,Scuti-like pulsations in the out-of-eclipse \tess\ light curve. As in the case of HD\,46052, the periodogram depicted in Fig.~\ref{fig:freq_HD46052_HD126031} (right upper panel) shows uniquely two peaks corresponding to $f_1$\,=\,0.268~d$^{-1}$ ($P$\,$\approx$\,3.73~d, i.e. nearly equal to the orbital period) and $f_2$\,=\,0.529~d$^{-1}$ (half orbital period $P$\,$\approx$\,1.891~d). No other power excess is present above the 99$\%$ confidence level, so we rule out that HD\,126031 may be a $\delta$\,Sct variable.

\begin{table}
    \caption{Pulsation frequencies of HD\,151604.}
    \label{Tab:freq_HD151604}
    \centering
    \begin{tabular}{ccc}
    \hline
    \hline
    \noalign{\smallskip}
    Mode &     Frequency  & Amplitude   \\
         & (day$^{-1}$)   &  (ppm)      \\
    \hline
    \noalign{\smallskip}
$f_{1}$  &  0.04988 & 301 \\
$f_{2}$  &  0.09577	& 299 \\
$f_{3}$  &  0.60856	& 268 \\
$f_{4}$  &  0.15763	& 238 \\
$f_{5}$  &  0.66044	& 225 \\
    \noalign{\smallskip}
    \hline
    \end{tabular}
\end{table}

\subsection{HD\,151604 (= V916\,Her)}
\label{hd151604}
The orbital dynamics of the Am star HD\,151604 have been investigated in depth by \citet{2007MNRAS.380.1064C}. Their preliminary analysis revealed a highly eccentric orbit ($e$\,=\,0.566) with a period of $P$\,=\,19.69858 days and a mass ratio approaching unity, indicative of the system's intriguing orbital characteristics. Our study fully supports these results. In a complementary study, \citet{smalley2014} conducted WASP photometry, focusing on the observational aspects of the system. Their findings unveiled an eclipse event occurring along the orbit, providing valuable observational insights into HD\,151604.

In corroboration with the orbital and observational data, our comprehensive chemical analysis affirms the Am-star classification for both components of this system. The characteristic chemical peculiarities associated with Am stars were evident, both stars show a deficiency in calcium and scandium, other than in light elements such as carbon and slight oxygen, while they have moderate overabundant iron peak and heavy elements ($\approx$~0.5~dex), except for barium for which we found an overabundance of $\approx$\,2~dex. The chemical compositions of the two stars are almost consistent with each other.

As for HD\,42954, the \tess\ light curves of this system show short-period variations out of the eclipses.  The cleaned periodogram corresponding to the data obtained in the two consecutive sectors 51 and 52 (22 April -- 13 June 2022) is shown in Fig.~\ref{fig:freq_HD151604}, where all the peaks higher than 1.5 times the 99\% confidence level are indicated with $f_{i}$ and their corresponding values are reported in Table~\ref{Tab:freq_HD151604}.  

We note that the two smallest frequencies, $f_1$ and $f_2$ correspond to the orbital period ($P_1\approx 20$ d) and its half. However, apart from $f_1$ and $f_2$,  
the periodogram shows an excess of power at frequencies of about 0.6~d$^{-1}$, which are all lower than 5~d$^{-1}$. According to \citet{2010ApJ...713L.192G}, HD\,151604 can be classified as a $\gamma$ Dor variable. For this object, we can hypothesize that the pulsating star could be component A, as only this component falls within the instability strip of the $\gamma$ Dor \citep{2003ApJ...593.1049W}, as seen in Fig.~\ref{hr}. However, given the width of the error bars, this result could be further revised and corrected.

\subsection{HD\,195020 (= MP\,Del)}\label{195020}
HD 195020 first came into focus as an Am peculiar star through an objective-prism survey by \citet{1970PASP...82..321B}. A pivotal study performed by \citet{2008MNRAS.390..958I} employed an advanced version of the WD program to conduct a thorough analysis of the $UBV$ LCs and RVs from this eclipsing binary system. This investigation led to precise determinations of the orbital parameters, as well as the masses, radii, and temperatures of the binary components.

In the case of this star, we could not use the \citet{2008MNRAS.390..958I} data together with ours, due to a difference in the velocity of the system's center of mass between their solution and ours of $\approx$\,5~km\,s$^{-1}$. As far as the other orbital parameters are concerned, however, their solution is perfectly consistent with ours. Regarding the difference in the velocity of the center of mass between the two datasets, which are clustered together, we can speculate that it could be attributed to the presence of a third massive body orbiting outside the AB pair with a period greater than 15 years. 

We also check for pulsations but no power excess is present above the 99$\%$ confidence level (see bottom panel of Fig~\ref{fig:freq_HD46052_HD126031}).

It is worth noting that our photometric and spectroscopic analyses mark the first attempt to characterize the components in detail, as far as our knowledge goes. According to our findings, components have mass of 1.90\,$\pm$\,0.05~M$_\odot$ and 1.42\,$\pm$\,0.03~M$_\odot$ and radii of 2.64\,$\pm$\,0.03 and 1.57\,$\pm$\,0.02~R$_\odot$, for primary and secondary components, respectively. Regarding the chemical composition, this system shows anomalous behavior regarding the distribution of abundances in the atmospheres of the two components, which are inconsistent with each other. In particular, both show typical signs of Am, i.e., underabundance of calcium and/or scandium (calcium is solar in component A). On the other hand, the other elements do not turn out to be consistent with each other; in the case of HD\,195020B we notice a general overabundance of both light and heavy elements (especially Ba), while in the case of HD\,195020A we have a general tendency to have solar or slightly below solar abundances.

\subsection{HD\,216429 (= V364\,Lac)}
In a recent paper, \citet{2015AJ....149..131E} employed new mass-luminosity, mass-radius, and mass-effective temperature calibrations to ascertain the physical parameters of the system, including temperatures ($T_{\textrm{eff}}^{\textrm{A}}$\,=\,8250~K and $T_{\textrm{eff}}^{\textrm{B}}$\,=\,8500~K), masses ($M_{\textrm{A}}$\,=\,2.334~M$_\odot$ and $M_{\textrm{B}}$\,=\,2.296~M$_\odot$), and radii ($R_{\textrm{A}}$\,=\,3.308~R$_\odot$ and $R_{\textrm{B}}$\,=\,2.986~R$_\odot$) for both components. Specifically, the corresponding parameters determined in our current study are consistent, within the experimental errors, with those derived by these authors (see Table~\ref{tab:param}).  

 The radial velocity curves obtained for HD\,216429 clearly show the deviation of velocities around $\phi \approx$~0.5 due to the Rossiter-McLaughlin effect \citep[RM,][]{1924ApJ....60...15R,1924ApJ....60...22M}. Unfortunately, the number of data at the phases where the effect is visible is not sufficient for an adequate sampling for the analysis. Nevertheless, since observations of the RM effect are rare in eclipse binary systems \citep{2007A&A...474..565A,1996PASP..108..962W,1982Ap&SS..81..357H}, it is important to highlight its presence in this star which can be the subject of future-focused studies.

As we already discussed in Sect~\ref{Subsec:photo}, this star shows a change in the separation between secondary and primary minima that spans from $0.5\cdot P_{\rm orb}$+0.063 days to $0.5\cdot P_{\rm orb}$+0.056 days between the two epochs of \tess\ observations (see Fig.~\ref{Fig:o-c}). This is clearly due to apsidal motion.

Interestingly, an earlier classification of this object as an uncertain peculiarity was presented by \citet{2009A&A...498..961R}. Our spectroscopic analysis revealed a similar composition in both components, compatible with that of the sun, although component A shows signs of a moderate Am-type peculiarity, namely a slight underabundance of calcium. An exception to this statement is oxygen, which in both stars shows an overabundance relative to the solar value of about 1.5 dex. 

\section{Discussion and conclusions}\label{sec:concl}

In this paper, we undertook photometric and spectroscopic analyses of six eclipsing binaries reported in the "General catalogue of Ap and Am stars" \citep{1991A&AS...89..429R,2009A&A...498..961R} as confirmed or suspected Am stars. Photometric data are collected from the \tess\ satellite, while time-resolved spectroscopy has been performed with the CAOS spectrograph at the Catania Astrophysical Observatory.

The solution of LC and RV curves allowed us to determine the orbital parameters of these systems and the stellar parameters of their components with good accuracy, especially as regards masses, radii, and temperature ratios. With the above stellar parameters, we have computed the luminosity of each component and positioned them on the HR diagram, enabling us to estimate their ages through isochrone fitting. It is important to note that the evolutionary track of each component is consistent, within the errors, with the dynamically derived mass. The accurate times of minima measured in the \tess\ LCs revealed apsidal motion in HD~216429, a system with eccentric orbit and for which the radial velocity curve clearly shows the Rossiter-McLaughlin effect around the conjunction orbital phase.

In the past, Am stars were believed to be unable to show pulsations because of processes such as diffusive settling and radiative levitation, generally attributed to chemical anomalies \citep{1970ApJ...160..641M,1983ApJ...269..239M}. However, despite this belief, many Am-type stars have been observed pulsating \citep{1989MNRAS.238.1077K,1999MNRAS.309..871M}. Recently, \citet{2021tsc2.confE..38G} analyzed \tess\ light curves of a sample of 55 Am stars located in the instability strips shown in Fig.~\ref{hr} and identified as a preliminary result two $\delta$~Sct stars (HD\,155316 and HD\,211643) and two $\delta$~Sct/$\gamma$~Dor hybrid candidates (HD\,8251 and HD\,108449). According to \citet{2017MNRAS.465.2662S}, excitation occurs mainly in the H/HeI ionization layer, driven by turbulent pressure.

With a Fourier analysis of \tess\ light curves, we found that HD\,42954 is a pulsating variable of type $\delta$~Sct and HD\,151604 shows pulsations of type $\gamma$~Dor. Given the position of the components within the instability strips for these classes of pulsating stars, we can infer that actually, the $\gamma$~Dor variable could be HD\,151604A, while no assumption can be made for HD\,42954. We also ruled out the presence of pulsations in HD\,46052, HD\,126031, and in HD\,195020. We did not look for pulsations for HD\,216429 since both components lie outside the blue edge of the $\delta$~Sct instability strip.

CAOS spectroscopy allowed us to derive the atmospheric parameters of the components of these systems and to study in detail their chemical abundances. We found that for four of them both components belong to the Am class, while for two of them, HD\,126031 and HD\,216429, only the main component is an Am star, whereas the secondaries show typically solar abundances. It is worth mentioning how in the case of the latter system, both stars show a conspicuous oxygen overabundance of about 1.5~dex more than the sun.

\citet{2023MNRAS.526.3386R} draw attention to the need to use chemical abundances based on non-LTE, especially for elements such as calcium and scandium, which are an important classification criterion for Am stars. For calcium in particular, their calculations are compatible with a positive correction of about 0.1-0.2~dex between LTE and non-LTE, even though the Am stars studied by these authors are significantly hotter than our targets. Thus, in general, the calculation of abundances in non-LTE, where possible, could lead to a better classification of these objects.

The age distribution found for our targets is in agreement with that of the Am stars presented in \citet{catanzaro2019,2022MNRAS.515.4350C}. On this topic, we discussed the possibility recently proposed by \citet{worthey2015}, of using Am stars as age indicators in stellar populations because of their extremely narrow age distribution. Our results, although consistent in the shape of the distribution,
shift the ages of the Am stars further back by about 0.35 Gyr than \citet{worthey2015}'s.
\begin{landscape}
\begin{table}
\caption{Orbital and astrophysical parameters derived in this study for our targets.}
\label{tab:param}
\centering
\begin{tabular}{lcccccc}
\hline
\hline
\noalign{\smallskip}
                                   & HD\,42954 & HD\,46052 & HD\,126031 & HD\,151604 & HD\,195020 & HD\,216429\\
\noalign{\smallskip}
\hline
\noalign{\smallskip}
       $T_0^a$                 &      \dots             & 8844.664157$\pm$0.000004    & 9665.508558$\pm$0.000008    &     \dots                 & 9790.47463$\pm$0.00005    & 8739.87082$\pm$0.00002 \\
       $T_{\rm per}^b$         & 9906.874 $\pm$ 0.005   & 9317.031 $\pm$ 0.001        & 9800.284  $\pm$ 0.001       & 9425.918 $\pm$ 0.003      & 9485.74 $\pm$ 0.02        & 9820.404 $\pm$ 0.002   \\
       $P_{\rm orb}$ (days)    & 23.80808 $\pm$ 0.00003 & 2.525019 $\pm$ 0.000001     & 3.782640 $\pm$ 0.000002     & 19.69846 $\pm$ 0.00001    & 21.33865 $\pm$ 0.00004    & 7.351486 $\pm$ 0.000002  \\       
       e                       & 0.55 $\pm$ 0.01        & 0.00 $\pm$ 0.01             & 0.00 $\pm$ 0.01             & 0.57 $\pm$ 0.01           & 0.22 $\pm$ 0.01           & 0.29 $\pm$ 0.01      \\
       $\omega (^\circ)$       & 266 $\pm$ 1            & \dots            & \dots              & 294 $\pm$ 1               & 291 $\pm$ 1               & 265 $\pm$ 1       \\
       $\gamma$ (km s$^{-1}$)  & 33.57 $\pm$ 0.07       & $-$9.1 $\pm$ 0.2            & $-$27.6 $\pm$ 0.4           & $-$14.26 $\pm$ 0.05       & $-$0.03 $\pm$ 0.02        & $-$11.3 $\pm$ 0.1   \\
       $K_{\textrm{A}}$ (km s$^{-1}$)     & 63.6 $\pm$ 0.1         & 117.6 $\pm$ 0.3             & 81.35 $\pm$ 0.06            & 71.4 $\pm$ 0.2            & 58.3 $\pm$ 0.7            & 94.5 $\pm$ 0.5      \\
       $K_{\textrm{B}}$ (km s$^{-1}$)     & 77.6 $\pm$ 0.6         & 126.2 $\pm$ 0.3             & 110.1 $\pm$ 0.1             & 72.3 $\pm$ 0.2            & 58.2 $\pm$ 0.9            & 96.1 $\pm$ 0.1      \\
       $M_{\rm{A}}$/$M_{\textrm{B}}$      & 1.22 $\pm$ 0.01        & 1.073 $\pm$ 0.005           & 1.35 $\pm$ 0.02             & 1.013 $\pm$ 0.004         & 1.00 $\pm$ 0.03           & 1.017 $\pm$ 0.007     \\
       $a_{\textrm{A}}$ (R$_\odot$)       & 25.0 $\pm$ 0.1$^*$     & 5.86 $\pm$ 0.01             & 6.12 $\pm$ 0.01             & 22.83 $\pm$ 0.07$^*$      & 20.64 $\pm$ 0.15          & 13.13 $\pm$ 0.08     \\
       $a_{\textrm{B}}$ (R$_\odot$)       & 30.5 $\pm$ 0.3$^*$     & 6.30 $\pm$ 0.03             & 8.29 $\pm$ 0.02             & 23.12 $\pm$ 0.07$^*$      & 27.63 $\pm$ 0.31          & 13.35 $\pm$ 0.03     \\
       {\it i}  (\degr)                   &     \dots              & 87.560 $\pm$ 0.002          & 83.22 $\pm$ 0.02            &     \dots                 & 88.280 $\pm$ 0.001        & 89.425 $\pm$ 0.006   \\
       $J^c$                              &  \dots                 &  0.9021 $\pm$ 0.0002        &  0.6711 $\pm$ 0.0005        & \dots                     & 0.7647 $\pm$ 0.0001       &  0.9518 $\pm$ 0.0003  \\
       $M_{\textrm{A}}$ (M$_\odot$)       &  2.06 $\pm$ 0.03$^*$   & 1.95 $\pm$ 0.01             & 1.631 $\pm$ 0.005           &   1.70 $\pm$ 0.01$^*$     & 1.90 $\pm$ 0.05           & 2.34 $\pm$ 0.01        \\
       $M_{\textrm{B}}$ (M$_\odot$)       &  1.69 $\pm$ 0.02$^*$   & 1.82 $\pm$ 0.01             & 1.208 $\pm$ 0.003           & 1.68 $\pm$ 0.01$^*$       & 1.42 $\pm$ 0.03           & 2.30 $\pm$ 0.02      \\
       $R_{\textrm{A}}$ (R$_\odot$)       &  4.2 $\pm$ 0.8$^{\dag}$ & 1.94 $\pm$ 0.02            & 1.94 $\pm$ 0.03             & 2.5 $\pm$ 0.2$^{\dag}$    & 2.64 $\pm$ 0.03           & 3.30 $\pm$ 0.11       \\
       $R_{\textrm{B}}$ (R$_\odot$)       &  3.2 $\pm$ 0.6$^{\dag}$ & 1.85 $\pm$ 0.02            & 1.25 $\pm$ 0.06             & 2.3 $\pm$ 0.2$^{\dag}$    & 1.57 $\pm$ 0.02           & 2.94 $\pm$ 0.12     \\
       log\,$g_{\textrm{A}}$              &    3.0$^{\dag}$        & 4.15 $\pm$ 0.01             & 4.07 $\pm$ 0.01             &   3.5$^{\dag}$            & 3.87 $\pm$ 0.02           & 3.77 $\pm$ 0.03       \\
       log\,$g_{\textrm{B}}$              &    3.5$^{\dag}$        & 4.16 $\pm$ 0.01             & 4.33 $\pm$ 0.04             &   3.5$^{\dag}$            & 4.20 $\pm$ 0.03           & 3.87 $\pm$ 0.04      \\
\noalign{\smallskip}
\hline
\noalign{\smallskip}
       $T_{\textrm {eff}}^{\textrm{A}}$ (K) & 7500 $\pm$ 300         & 8750 $\pm$ 200              & 7500 $\pm$ 150              & 7000 $\pm$ 300            & 7750 $\pm$ 200            & 8500 $\pm$ 300        \\
       $T_{\textrm {eff}}^{\textrm{B}}$ (K) & 8000 $\pm$ 500         & 8250 $\pm$ 200              & 6450 $\pm$ 200              & 7250 $\pm$ 300            & 7000 $\pm$ 250            & 8750 $\pm$ 300        \\
       $v_{\textrm {A}}$\,sin\,$i$ (\kms)   & 15 $\pm$ 1             & 37 $\pm$ 4                  & 28 $\pm$ 3                  & 10 $\pm$ 1                & 25 $\pm$ 3                & 45 $\pm$ 4           \\
       $v_{\textrm {B}}$\,sin\,$i$ (\kms)   & 15 $\pm$ 1             & 37 $\pm$ 4                  & 17 $\pm$ 2                  & 10 $\pm$ 1                & 40 $\pm$ 4                & 15 $\pm$ 1           \\
       $\xi_{\textrm {A}}$                  & 3.0 $\pm$ 0.3          & 3.1 $\pm$ 0.3               & 2.7 $\pm$ 0.3               & 2.8 $\pm$ 0.5             & 2.7 $\pm$ 0.8             & 3.1 $\pm$ 0.5        \\
       $\xi_{\textrm {B}}$                  & 3.3 $\pm$ 0.3          & 3.3 $\pm$ 0.2               & 1.8 $\pm$ 0.8               & 2.3 $\pm$ 0.6             & 3.0 $\pm$ 0.2             & 2.9 $\pm$ 0.6        \\
       $l_{H\alpha}$                        &   \dots                & 1.29 $\pm$ 0.05             & 4.8 $\pm$ 0.5               &     \dots                 & 3.9 $\pm$ 0.2             & 1.2 $\pm$ 0.3        \\
       $l_{H\beta}$                         &   \dots                & 1.37 $\pm$ 0.06             & 6.0 $\pm$ 0.7               &     \dots                 & 4.4 $\pm$ 0.2             & 1.1 $\pm$ 0.3         \\
       $l_\lambda$                          & 1.4 $\pm$ 0.4          &  \dots                      & \dots                       & 1.2 $\pm$ 0.5             & \dots                     & \dots                 \\
       $L_{\textrm {A}}$ (L$_\odot$)        &  49.8 $\pm$ 18.0       & 19.8 $\pm$ 1.8              & 10.7 $\pm$ 0.9              & 13.4 $\pm$ 2.3            & 22.5 $\pm$ 1.3            & 51.0 $\pm$ 8.0      \\
       $L_{\textrm {B}}$ (L$_\odot$)        &  36.7 $\pm$ 14.2       & 14.1 $\pm$ 1.4              &  2.4 $\pm$ 0.4              & 13.3 $\pm$ 2.3            &  5.3 $\pm$ 1.4            & 45.5 $\pm$ 7.3      \\
       age (Gyr)                            &  0.7 $\pm$ 0.1         & 0.6 $\pm$ 0.1               & 1.1 $\pm$ 0.2               & 1.3 $\pm$ 0.3             & 0.93 $\pm$ 0.04           & 0.6 $\pm$ 0.1     \\ 
\noalign{\smallskip}
\hline
\noalign{\smallskip}
\end{tabular}

{\it Note.} $^a$\,Time of primary minimum (HJD$-2\,450\,000)$. 
$^b$\,Time of periastron passage (HJD$-2\,450\,000)$.
$^c$ Surface brightness ratio 
$^*$ Dynamical semi-major axes and mass ($a\sin i$ and $M\sin^3i$). 
$^{\dag}$ From the spectral synthesis.
\end{table}
\end{landscape}

\begin{acknowledgements}
{\bf We thank our anonymous Referee for her/his helpful comments.}
This research made use of SIMBAD and VIZIER databases, operated at the CDS, Strasbourg, France. 
This paper includes data collected by the \tess\ mission which are publicly available from the Mikulski Archive for Space Telescopes (MAST). 
Funding for the \tess\ mission is provided by the NASA's Science Mission Directorate.
This work has made use of data from the European Space Agency (ESA)
mission {\it Gaia} ({\tt https://www.cosmos.esa.int/gaia}), processed by
the {\it Gaia} Data Processing and Analysis Consortium (DPAC,
{\tt https://www.cosmos.esa.int/web/gaia/dpac/consortium}). Funding
for the DPAC has been provided by national institutions, in particular,
the institutions participating in the {\it Gaia} Multilateral Agreement.

AF acknowledges funding from the Large-Grant INAF YODA (YSOs Outflow, Disks and Accretion). 
\end{acknowledgements}

\bibliographystyle{aa}
\bibliography{SB2_Caos}

\begin{thebibliography}{69}
\expandafter\ifx\csname natexlab\endcsname\relax\def\natexlab#1{#1}\fi

\bibitem[{{Abt} \& {Levy}(1985)}]{abt85}
{Abt}, H.~A. \& {Levy}, S.~G. 1985, \apjs, 59, 229

\bibitem[{{Albrecht} {et~al.}(2007){Albrecht}, {Reffert}, {Snellen},
  {Quirrenbach}, \& {Mitchell}}]{2007A&A...474..565A}
{Albrecht}, S., {Reffert}, S., {Snellen}, I., {Quirrenbach}, A., \& {Mitchell},
  D.~S. 2007, \aap, 474, 565

\bibitem[{{Allard} {et~al.}(2012){Allard}, {Homeier}, \&
  {Freytag}}]{Allard2012}
{Allard}, F., {Homeier}, D., \& {Freytag}, B. 2012, Philosophical Transactions
  of the Royal Society of London Series A, 370, 2765

\bibitem[{{Alonso-Santiago} {et~al.}(2023){Alonso-Santiago}, {Frasca},
  {Catanzaro}, {Bragaglia}, {Magrini}, {Vallenari}, {Carretta}, \&
  {Lucatello}}]{2023arXiv231208581A}
{Alonso-Santiago}, J., {Frasca}, A., {Catanzaro}, G., {et~al.} 2023, arXiv
  e-prints, arXiv:2312.08581

\bibitem[{{Baroch} {et~al.}(2021){Baroch}, {Gim{\'e}nez}, {Ribas}, {Morales},
  {Anglada-Escud{\'e}}, \& {Claret}}]{Baroch2021}
{Baroch}, D., {Gim{\'e}nez}, A., {Ribas}, I., {et~al.} 2021, \aap, 649, A64

\bibitem[{{Bidelman}(1988)}]{1988PASP..100.1084B}
{Bidelman}, W.~P. 1988, \pasp, 100, 1084

\bibitem[{{Bond}(1970)}]{1970PASP...82..321B}
{Bond}, H.~E. 1970, \pasp, 82, 321

\bibitem[{{Breger} \& {Pamyatnykh}(1998)}]{1998A&A...332..958B}
{Breger}, M. \& {Pamyatnykh}, A.~A. 1998, \aap, 332, 958

\bibitem[{{Bressan} {et~al.}(2012){Bressan}, {Marigo}, {Girardi}, {Salasnich},
  {Dal Cero}, {Rubele}, \& {Nanni}}]{2012MNRAS.427..127B}
{Bressan}, A., {Marigo}, P., {Girardi}, L., {et~al.} 2012, \mnras, 427, 127

\bibitem[{{Caldwell} {et~al.}(2020){Caldwell}, {Tenenbaum}, {Twicken},
  {Jenkins}, {Ting}, {Smith}, {Hedges}, {Fausnaugh}, {Rose}, \&
  {Burke}}]{Caldwell2020}
{Caldwell}, D.~A., {Tenenbaum}, P., {Twicken}, J.~D., {et~al.} 2020, Research
  Notes of the American Astronomical Society, 4, 201

\bibitem[{{Carquillat} \& {Prieur}(2007)}]{2007MNRAS.380.1064C}
{Carquillat}, J.~M. \& {Prieur}, J.~L. 2007, \mnras, 380, 1064

\bibitem[{{Carquillat} {et~al.}(2004){Carquillat}, {Prieur}, {Ginestet},
  {Oblak}, \& {Kurpinska-Winiarska}}]{2004MNRAS.352..708C}
{Carquillat}, J.~M., {Prieur}, J.~L., {Ginestet}, N., {Oblak}, E., \&
  {Kurpinska-Winiarska}, M. 2004, \mnras, 352, 708

\bibitem[{{Castelli} \& {Hubrig}(2004)}]{2004A&A...425..263C}
{Castelli}, F. \& {Hubrig}, S. 2004, \aap, 425, 263

\bibitem[{{Catanzaro}(2006)}]{catanzaro2006}
{Catanzaro}, G. 2006, \mnras, 368, 247

\bibitem[{{Catanzaro} {et~al.}(2019){Catanzaro}, {Bus{\`a}}, {Gangi},
  {Giarrusso}, {Leone}, \& {Munari}}]{catanzaro2019}
{Catanzaro}, G., {Bus{\`a}}, I., {Gangi}, M., {et~al.} 2019, \mnras, 484, 2530

\bibitem[{{Catanzaro} {et~al.}(2022){Catanzaro}, {Colombo}, {Ferrara}, \&
  {Giarrusso}}]{2022MNRAS.515.4350C}
{Catanzaro}, G., {Colombo}, C., {Ferrara}, C., \& {Giarrusso}, M. 2022, \mnras,
  515, 4350

\bibitem[{{Catanzaro} {et~al.}(2016){Catanzaro}, {Giarrusso}, {Leone},
  {Munari}, {Scalia}, {Sparacello}, \& {Scuderi}}]{2016MNRAS.460.1999C}
{Catanzaro}, G., {Giarrusso}, M., {Leone}, F., {et~al.} 2016, \mnras, 460, 1999

\bibitem[{{Catanzaro} {et~al.}(2020){Catanzaro}, {Giarrusso}, {Munari}, \&
  {Leone}}]{2020MNRAS.499.3720C}
{Catanzaro}, G., {Giarrusso}, M., {Munari}, M., \& {Leone}, F. 2020, \mnras,
  499, 3720

\bibitem[{{Catanzaro} {et~al.}(2011){Catanzaro}, {Ripepi}, {Bernabei},
  {Marconi}, {Balona}, {Kurtz}, {Smalley}, {Borucki}, {Bruntt},
  {Christensen-Dalsgaard}, {Grigahc{\`e}ne}, {Kjeldsen}, {Koch}, {Monteiro},
  {Su{\'a}rez}, {Szab{\'o}}, \& {Uytterhoeven}}]{2011MNRAS.411.1167C}
{Catanzaro}, G., {Ripepi}, V., {Bernabei}, S., {et~al.} 2011, \mnras, 411, 1167

\bibitem[{{Catanzaro} {et~al.}(2015){Catanzaro}, {Ripepi}, {Biazzo},
  {Bus{\'a}}, {Frasca}, {Leone}, {Giarrusso}, {Munari}, \&
  {Scuderi}}]{catanzaro2015}
{Catanzaro}, G., {Ripepi}, V., {Biazzo}, K., {et~al.} 2015, \mnras, 451, 184

\bibitem[{{Catanzaro} {et~al.}(2013){Catanzaro}, {Ripepi}, \&
  {Bruntt}}]{catanzaro2013}
{Catanzaro}, G., {Ripepi}, V., \& {Bruntt}, H. 2013, \mnras, 431, 3258

\bibitem[{{Claret}(2017)}]{Claret2017}
{Claret}, A. 2017, \aap, 600, A30

\bibitem[{{Cowley} {et~al.}(1969){Cowley}, {Cowley}, {Jaschek}, \&
  {Jaschek}}]{cowley69}
{Cowley}, A., {Cowley}, C., {Jaschek}, M., \& {Jaschek}, C. 1969, \aj, 74, 375

\bibitem[{{Eker} {et~al.}(2015){Eker}, {Soydugan}, {Soydugan}, {Bilir}, {Yaz
  G{\"o}k{\c{c}}e}, {Steer}, {T{\"u}ys{\"u}z}, {{\c{S}}eny{\"u}z}, \&
  {Demircan}}]{2015AJ....149..131E}
{Eker}, Z., {Soydugan}, F., {Soydugan}, E., {et~al.} 2015, \aj, 149, 131

\bibitem[{{Frasca} {et~al.}(2023){Frasca}, {Alonso-Santiago}, {Catanzaro},
  {Bragaglia}, {D'Orazi}, {Fu}, {Vallenari}, \& {Andreuzzi}}]{Frasca2023}
{Frasca}, A., {Alonso-Santiago}, J., {Catanzaro}, G., {et~al.} 2023, \aap, 677,
  A154

\bibitem[{{Frasca} {et~al.}(2022){Frasca}, {Catanzaro}, {Bus{\`a}}, {Guillout},
  {Alonso-Santiago}, {Ferrara}, {Giarrusso}, {Munari}, \& {Leone}}]{Frasca2022}
{Frasca}, A., {Catanzaro}, G., {Bus{\`a}}, I., {et~al.} 2022, \mnras, 515, 3716

\bibitem[{{Fu} {et~al.}(2020){Fu}, {Cat}, {Zong}, {Frasca}, {Gray}, {Ren},
  {Molenda-{\.Z}akowicz}, {Corbally}, {Catanzaro}, {Shi}, {Luo}, \&
  {Zhang}}]{2020RAA....20..167F}
{Fu}, J.-N., {Cat}, P.~D., {Zong}, W., {et~al.} 2020, Research in Astronomy and
  Astrophysics, 20, 167

\bibitem[{{Gaia Collaboration} {et~al.}(2023){Gaia Collaboration}, {Vallenari},
  {Brown}, {Prusti}, {de Bruijne}, {Arenou}, {Babusiaux}, {Biermann},
  {Creevey}, {Ducourant}, {Evans}, {Eyer}, {Guerra}, {Hutton}, {Jordi},
  {Klioner}, {Lammers}, {Lindegren}, {Luri}, {Mignard}, {Panem}, {Pourbaix},
  {Randich}, {Sartoretti}, {Soubiran}, {Tanga}, {Walton}, {Bailer-Jones},
  {Bastian}, {Drimmel}, {Jansen}, {Katz}, {Lattanzi}, {van Leeuwen}, {Bakker},
  {Cacciari}, {Casta{\~n}eda}, {De Angeli}, {Fabricius}, {Fouesneau},
  {Fr{\'e}mat}, {Galluccio}, {Guerrier}, {Heiter}, {Masana}, {Messineo},
  {Mowlavi}, {Nicolas}, {Nienartowicz}, {Pailler}, {Panuzzo}, {Riclet}, {Roux},
  {Seabroke}, {Sordo}, {Th{\'e}venin}, {Gracia-Abril}, {Portell}, {Teyssier},
  {Altmann}, {Andrae}, {Audard}, {Bellas-Velidis}, {Benson}, {Berthier},
  {Blomme}, {Burgess}, {Busonero}, {Busso}, {C{\'a}novas}, {Carry}, {Cellino},
  {Cheek}, {Clementini}, {Damerdji}, {Davidson}, {de Teodoro}, {Nu{\~n}ez
  Campos}, {Delchambre}, {Dell'Oro}, {Esquej}, {Fern{\'a}ndez-Hern{\'a}ndez},
  {Fraile}, {Garabato}, {Garc{\'\i}a-Lario}, {Gosset}, {Haigron}, {Halbwachs},
  {Hambly}, {Harrison}, {Hern{\'a}ndez}, {Hestroffer}, {Hodgkin}, {Holl},
  {Jan{\ss}en}, {Jevardat de Fombelle}, {Jordan}, {Krone-Martins}, {Lanzafame},
  {L{\"o}ffler}, {Marchal}, {Marrese}, {Moitinho}, {Muinonen}, {Osborne},
  {Pancino}, {Pauwels}, {Recio-Blanco}, {Reyl{\'e}}, {Riello}, {Rimoldini},
  {Roegiers}, {Rybizki}, {Sarro}, {Siopis}, {Smith}, {Sozzetti}, {Utrilla},
  {van Leeuwen}, {Abbas}, {{\'A}brah{\'a}m}, {Abreu Aramburu}, {Aerts},
  {Aguado}, {Ajaj}, {Aldea-Montero}, {Altavilla}, {{\'A}lvarez}, {Alves},
  {Anders}, {Anderson}, {Anglada Varela}, {Antoja}, {Baines}, {Baker},
  {Balaguer-N{\'u}{\~n}ez}, {Balbinot}, {Balog}, {Barache}, {Barbato},
  {Barros}, {Barstow}, {Bartolom{\'e}}, {Bassilana}, {Bauchet}, {Becciani},
  {Bellazzini}, {Berihuete}, {Bernet}, {Bertone}, {Bianchi}, {Binnenfeld},
  {Blanco-Cuaresma}, {Blazere}, {Boch}, {Bombrun}, {Bossini}, {Bouquillon},
  {Bragaglia}, {Bramante}, {Breedt}, {Bressan}, {Brouillet}, {Brugaletta},
  {Bucciarelli}, {Burlacu}, {Butkevich}, {Buzzi}, {Caffau}, {Cancelliere},
  {Cantat-Gaudin}, {Carballo}, {Carlucci}, {Carnerero}, {Carrasco},
  {Casamiquela}, {Castellani}, {Castro-Ginard}, {Chaoul}, {Charlot}, {Chemin},
  {Chiaramida}, {Chiavassa}, {Chornay}, {Comoretto}, {Contursi}, {Cooper},
  {Cornez}, {Cowell}, {Crifo}, {Cropper}, {Crosta}, {Crowley}, {Dafonte},
  {Dapergolas}, {David}, {David}, {de Laverny}, {De Luise}, {De March}, {De
  Ridder}, {de Souza}, {de Torres}, {del Peloso}, {del Pozo}, {Delbo},
  {Delgado}, {Delisle}, {Demouchy}, {Dharmawardena}, {Di Matteo}, {Diakite},
  {Diener}, {Distefano}, {Dolding}, {Edvardsson}, {Enke}, {Fabre}, {Fabrizio},
  {Faigler}, {Fedorets}, {Fernique}, {Fienga}, {Figueras}, {Fournier},
  {Fouron}, {Fragkoudi}, {Gai}, {Garcia-Gutierrez}, {Garcia-Reinaldos},
  {Garc{\'\i}a-Torres}, {Garofalo}, {Gavel}, {Gavras}, {Gerlach}, {Geyer},
  {Giacobbe}, {Gilmore}, {Girona}, {Giuffrida}, {Gomel}, {Gomez},
  {Gonz{\'a}lez-N{\'u}{\~n}ez}, {Gonz{\'a}lez-Santamar{\'\i}a},
  {Gonz{\'a}lez-Vidal}, {Granvik}, {Guillout}, {Guiraud},
  {Guti{\'e}rrez-S{\'a}nchez}, {Guy}, {Hatzidimitriou}, {Hauser}, {Haywood},
  {Helmer}, {Helmi}, {Sarmiento}, {Hidalgo}, {Hilger}, {H{\l}adczuk}, {Hobbs},
  {Holland}, {Huckle}, {Jardine}, {Jasniewicz}, {Jean-Antoine Piccolo},
  {Jim{\'e}nez-Arranz}, {Jorissen}, {Juaristi Campillo}, {Julbe}, {Karbevska},
  {Kervella}, {Khanna}, {Kontizas}, {Kordopatis}, {Korn}, {K{\'o}sp{\'a}l},
  {Kostrzewa-Rutkowska}, {Kruszy{\'n}ska}, {Kun}, {Laizeau}, {Lambert},
  {Lanza}, {Lasne}, {Le Campion}, {Lebreton}, {Lebzelter}, {Leccia}, {Leclerc},
  {Lecoeur-Taibi}, {Liao}, {Licata}, {Lindstr{\o}m}, {Lister}, {Livanou},
  {Lobel}, {Lorca}, {Loup}, {Madrero Pardo}, {Magdaleno Romeo}, {Managau},
  {Mann}, {Manteiga}, {Marchant}, {Marconi}, {Marcos}, {Marcos Santos},
  {Mar{\'\i}n Pina}, {Marinoni}, {Marocco}, {Marshall}, {Martin Polo},
  {Mart{\'\i}n-Fleitas}, {Marton}, {Mary}, {Masip}, {Massari},
  {Mastrobuono-Battisti}, {Mazeh}, {McMillan}, {Messina}, {Michalik}, {Millar},
  {Mints}, {Molina}, {Molinaro}, {Moln{\'a}r}, {Monari}, {Mongui{\'o}},
  {Montegriffo}, {Montero}, {Mor}, {Mora}, {Morbidelli}, {Morel}, {Morris},
  {Muraveva}, {Murphy}, {Musella}, {Nagy}, {Noval}, {Oca{\~n}a}, {Ogden},
  {Ordenovic}, {Osinde}, {Pagani}, {Pagano}, {Palaversa}, {Palicio},
  {Pallas-Quintela}, {Panahi}, {Payne-Wardenaar}, {Pe{\~n}alosa Esteller},
  {Penttil{\"a}}, {Pichon}, {Piersimoni}, {Pineau}, {Plachy}, {Plum}, {Poggio},
  {Pr{\v{s}}a}, {Pulone}, {Racero}, {Ragaini}, {Rainer}, {Raiteri}, {Rambaux},
  {Ramos}, {Ramos-Lerate}, {Re Fiorentin}, {Regibo}, {Richards}, {Rios Diaz},
  {Ripepi}, {Riva}, {Rix}, {Rixon}, {Robichon}, {Robin}, {Robin}, {Roelens},
  {Rogues}, {Rohrbasser}, {Romero-G{\'o}mez}, {Rowell}, {Royer}, {Ruz Mieres},
  {Rybicki}, {Sadowski}, {S{\'a}ez N{\'u}{\~n}ez}, {Sagrist{\`a} Sell{\'e}s},
  {Sahlmann}, {Salguero}, {Samaras}, {Sanchez Gimenez}, {Sanna},
  {Santove{\~n}a}, {Sarasso}, {Schultheis}, {Sciacca}, {Segol}, {Segovia},
  {S{\'e}gransan}, {Semeux}, {Shahaf}, {Siddiqui}, {Siebert}, {Siltala},
  {Silvelo}, {Slezak}, {Slezak}, {Smart}, {Snaith}, {Solano}, {Solitro},
  {Souami}, {Souchay}, {Spagna}, {Spina}, {Spoto}, {Steele},
  {Steidelm{\"u}ller}, {Stephenson}, {S{\"u}veges}, {Surdej}, {Szabados},
  {Szegedi-Elek}, {Taris}, {Taylor}, {Teixeira}, {Tolomei}, {Tonello}, {Torra},
  {Torra}, {Torralba Elipe}, {Trabucchi}, {Tsounis}, {Turon}, {Ulla}, {Unger},
  {Vaillant}, {van Dillen}, {van Reeven}, {Vanel}, {Vecchiato}, {Viala},
  {Vicente}, {Voutsinas}, {Weiler}, {Wevers}, {Wyrzykowski}, {Yoldas}, {Yvard},
  {Zhao}, {Zorec}, {Zucker}, \& {Zwitter}}]{GaiaDR3}
{Gaia Collaboration}, {Vallenari}, A., {Brown}, A.~G.~A., {et~al.} 2023, \aap,
  674, A1

\bibitem[{{Gebran} {et~al.}(2014){Gebran}, {Monier}, {Royer}, {Lobel}, \&
  {Blomme}}]{2014psce.conf..193G}
{Gebran}, M., {Monier}, R., {Royer}, F., {Lobel}, A., \& {Blomme}, R. 2014, in
  {Putting A Stars into Context: Evolution, Environment, and Related Stars},
  ed. G.~{Mathys}, E.~R. {Griffin}, O.~{Kochukhov}, R.~{Monier}, \& G.~M.
  {Wahlgren}, 193--198

\bibitem[{{Grenier} {et~al.}(1999){Grenier}, {Baylac}, {Rolland}, {Burnage},
  {Arenou}, {Briot}, {Delmas}, {Duflot}, {Genty}, {G{\'o}mez}, {Halbwachs},
  {Marouard}, {Oblak}, \& {Sellier}}]{1999A&AS..137..451G}
{Grenier}, S., {Baylac}, M.~O., {Rolland}, L., {et~al.} 1999, \aaps, 137, 451

\bibitem[{{Grevesse} {et~al.}(2010){Grevesse}, {Asplund}, {Sauval}, \&
  {Scott}}]{2010Ap&SS.328..179G}
{Grevesse}, N., {Asplund}, M., {Sauval}, A.~J., \& {Scott}, P. 2010, \apss,
  328, 179

\bibitem[{{Grigahc{\`e}ne} {et~al.}(2010){Grigahc{\`e}ne}, {Antoci}, {Balona},
  {Catanzaro}, {Daszy{\'n}ska-Daszkiewicz}, {Guzik}, {Handler}, {Houdek},
  {Kurtz}, {Marconi}, {Monteiro}, {Moya}, {Ripepi}, {Su{\'a}rez},
  {Uytterhoeven}, {Borucki}, {Brown}, {Christensen-Dalsgaard}, {Gilliland},
  {Jenkins}, {Kjeldsen}, {Koch}, {Bernabei}, {Bradley}, {Breger}, {Di
  Criscienzo}, {Dupret}, {Garc{\'\i}a}, {Garc{\'\i}a Hern{\'a}ndez},
  {Jackiewicz}, {Kaiser}, {Lehmann}, {Mart{\'\i}n-Ruiz}, {Mathias},
  {Molenda-{\.Z}akowicz}, {Nemec}, {Nuspl}, {Papar{\'o}}, {Roth}, {Szab{\'o}},
  {Suran}, \& {Ventura}}]{2010ApJ...713L.192G}
{Grigahc{\`e}ne}, A., {Antoci}, V., {Balona}, L., {et~al.} 2010, \apjl, 713,
  L192

\bibitem[{{Guzik} {et~al.}(2021){Guzik}, {Jackiewicz}, {Catanzaro}, \&
  {Soukup}}]{2021tsc2.confE..38G}
{Guzik}, J.~A., {Jackiewicz}, J., {Catanzaro}, G., \& {Soukup}, M.~S. 2021, in
  Posters from the TESS Science Conference II (TSC2), 38

\bibitem[{{Hube} \& {Couch}(1982)}]{1982Ap&SS..81..357H}
{Hube}, D.~P. \& {Couch}, J.~S. 1982, \apss, 81, 357

\bibitem[{{Ibano{\v{g}}lu} {et~al.}(2008){Ibano{\v{g}}lu}, {Evren},
  {Ta{\c{s}}}, {{\c{C}}ak{\i}rl{\i}}, {Bozkurt}, {Af{\c{s}}ar}, {Sipahi},
  {Dal}, {{\"O}zdarcan}, {{\c{C}}amurdan}, {{\c{C}}amurdan}, \&
  {Frasca}}]{2008MNRAS.390..958I}
{Ibano{\v{g}}lu}, C., {Evren}, S., {Ta{\c{s}}}, G., {et~al.} 2008, \mnras, 390,
  958

\bibitem[{{IJspeert} {et~al.}(2021){IJspeert}, {Tkachenko}, {Johnston},
  {Garcia}, {De Ridder}, {Van Reeth}, \& {Aerts}}]{2021A&A...652A.120I}
{IJspeert}, L.~W., {Tkachenko}, A., {Johnston}, C., {et~al.} 2021, \aap, 652,
  A120

\bibitem[{{Kahraman Ali{\c{c}}avu{\c{s}}} \&
  {Ali{\c{c}}avu{\c{s}}}(2020)}]{filiz2020}
{Kahraman Ali{\c{c}}avu{\c{s}}}, F. \& {Ali{\c{c}}avu{\c{s}}}, F. 2020,
  Research in Astronomy and Astrophysics, 20, 150

\bibitem[{{Kurtz}(1989)}]{1989MNRAS.238.1077K}
{Kurtz}, D.~W. 1989, \mnras, 238, 1077

\bibitem[{{Kurucz}(1993)}]{1993ASPC...44...87K}
{Kurucz}, R.~L. 1993, in {Astronomical Society of the Pacific Conference
  Series}, Vol.~44, {IAU Colloq. 138: Peculiar versus Normal Phenomena in
  A-type and Related Stars}, ed. M.~M. {Dworetsky}, F.~{Castelli}, \&
  R.~{Faraggiana}, 87

\bibitem[{{Kurucz} \& {Avrett}(1981)}]{1981SAOSR.391.....K}
{Kurucz}, R.~L. \& {Avrett}, E.~H. 1981, SAO Special Report, 391

\bibitem[{{Leone} {et~al.}(2016){Leone}, {Avila}, {Bellassai}, {Bruno},
  {Catalano}, {Di Benedetto}, {Di Stefano}, {Gangi}, {Giarrusso}, {Greco},
  {Martinetti}, {Miraglia}, {Munari}, {Pontoni}, {Scalia}, {Scuderi}, \&
  {Span{\'o}}}]{2016AJ....151..116L}
{Leone}, F., {Avila}, G., {Bellassai}, G., {et~al.} 2016, \aj, 151, 116

\bibitem[{{Leone} \& {Catanzaro}(1998)}]{1998A&A...331..627L}
{Leone}, F. \& {Catanzaro}, G. 1998, \aap, 331, 627

\bibitem[{{Martinez} {et~al.}(1999){Martinez}, {Kurtz}, {Ashoka}, {Chaubey},
  {Gupta}, {Leone}, {Catanzaro}, {Sagar}, {Raj}, {Seetha}, \&
  {Kasturirangan}}]{1999MNRAS.309..871M}
{Martinez}, P., {Kurtz}, D.~W., {Ashoka}, B.~N., {et~al.} 1999, \mnras, 309,
  871

\bibitem[{{Matson} {et~al.}(2019){Matson}, {Howell}, \&
  {Ciardi}}]{2019AJ....157..211M}
{Matson}, R.~A., {Howell}, S.~B., \& {Ciardi}, D.~R. 2019, \aj, 157, 211

\bibitem[{{McLaughlin}(1924)}]{1924ApJ....60...22M}
{McLaughlin}, D.~B. 1924, \apj, 60, 22

\bibitem[{{Michaud}(1970)}]{1970ApJ...160..641M}
{Michaud}, G. 1970, \apj, 160, 641

\bibitem[{{Michaud} {et~al.}(1983){Michaud}, {Tarasick}, {Charland}, \&
  {Pelletier}}]{1983ApJ...269..239M}
{Michaud}, G., {Tarasick}, D., {Charland}, Y., \& {Pelletier}, C. 1983, \apj,
  269, 239

\bibitem[{{Nelson} \& {Davis}(1972)}]{NelsonDavis1972}
{Nelson}, B. \& {Davis}, W.~D. 1972, \apj, 174, 617

\bibitem[{{Nguyen} {et~al.}(2022){Nguyen}, {Costa}, {Girardi}, {Volpato},
  {Bressan}, {Chen}, {Marigo}, {Fu}, \& {Goudfrooij}}]{2022A&A...665A.126N}
{Nguyen}, C.~T., {Costa}, G., {Girardi}, L., {et~al.} 2022, \aap, 665, A126

\bibitem[{{Pojmanski}(2002)}]{2002AcA....52..397P}
{Pojmanski}, G. 2002, \actaa, 52, 397

\bibitem[{{Renson} {et~al.}(1991){Renson}, {Gerbaldi}, \&
  {Catalano}}]{1991A&AS...89..429R}
{Renson}, P., {Gerbaldi}, M., \& {Catalano}, F.~A. 1991, \aaps, 89, 429

\bibitem[{{Renson} \& {Manfroid}(2009)}]{2009A&A...498..961R}
{Renson}, P. \& {Manfroid}, J. 2009, \aap, 498, 961

\bibitem[{{Ricker} {et~al.}(2015){Ricker}, {Winn}, {Vanderspek}, {Latham},
  {Bakos}, {Bean}, {Berta-Thompson}, {Brown}, {Buchhave}, {Butler}, {Butler},
  {Chaplin}, {Charbonneau}, {Christensen-Dalsgaard}, {Clampin}, {Deming},
  {Doty}, {De Lee}, {Dressing}, {Dunham}, {Endl}, {Fressin}, {Ge}, {Henning},
  {Holman}, {Howard}, {Ida}, {Jenkins}, {Jernigan}, {Johnson}, {Kaltenegger},
  {Kawai}, {Kjeldsen}, {Laughlin}, {Levine}, {Lin}, {Lissauer}, {MacQueen},
  {Marcy}, {McCullough}, {Morton}, {Narita}, {Paegert}, {Palle}, {Pepe},
  {Pepper}, {Quirrenbach}, {Rinehart}, {Sasselov}, {Sato}, {Seager},
  {Sozzetti}, {Stassun}, {Sullivan}, {Szentgyorgyi}, {Torres}, {Udry}, \&
  {Villasenor}}]{Ricker2015}
{Ricker}, G.~R., {Winn}, J.~N., {Vanderspek}, R., {et~al.} 2015, Journal of
  Astronomical Telescopes, Instruments, and Systems, 1, 014003

\bibitem[{{Roberts} {et~al.}(1987){Roberts}, {Lehar}, \&
  {Dreher}}]{Roberts1987}
{Roberts}, D.~H., {Lehar}, J., \& {Dreher}, J.~W. 1987, \aj, 93, 968

\bibitem[{{Romanovskaya} {et~al.}(2023){Romanovskaya}, {Ryabchikova},
  {Pakhomov}, {Korotin}, \& {Sitnova}}]{2023MNRAS.526.3386R}
{Romanovskaya}, A.~M., {Ryabchikova}, T.~A., {Pakhomov}, Y.~V., {Korotin},
  S.~A., \& {Sitnova}, T.~M. 2023, \mnras, 526, 3386

\bibitem[{{Rossiter}(1924)}]{1924ApJ....60...15R}
{Rossiter}, R.~A. 1924, \apj, 60, 15

\bibitem[{{Scargle}(1982)}]{Scargle1982}
{Scargle}, J.~D. 1982, \apj, 263, 835

\bibitem[{{Smalley} {et~al.}(2017){Smalley}, {Antoci}, {Holdsworth}, {Kurtz},
  {Murphy}, {De Cat}, {Anderson}, {Catanzaro}, {Collier Cameron}, {Hellier},
  {Maxted}, {Norton}, {Pollacco}, {Ripepi}, {West}, \&
  {Wheatley}}]{2017MNRAS.465.2662S}
{Smalley}, B., {Antoci}, V., {Holdsworth}, D.~L., {et~al.} 2017, \mnras, 465,
  2662

\bibitem[{{Smalley} {et~al.}(2014){Smalley}, {Southworth}, {Pintado}, {Gillon},
  {Holdsworth}, {Anderson}, {Barros}, {Collier Cameron}, {Delrez}, {Faedi},
  {Haswell}, {Hellier}, {Horne}, {Jehin}, {Maxted}, {Norton}, {Pollacco},
  {Skillen}, {Smith}, {West}, \& {Wheatley}}]{smalley2014}
{Smalley}, B., {Southworth}, J., {Pintado}, O.~I., {et~al.} 2014, \aap, 564,
  A69

\bibitem[{{Southworth}(2013)}]{Southworth2013}
{Southworth}, J. 2013, \aap, 557, A119

\bibitem[{{Southworth} {et~al.}(2004){Southworth}, {Maxted}, \&
  {Smalley}}]{Southworth2004MNRAS}
{Southworth}, J., {Maxted}, P.~F.~L., \& {Smalley}, B. 2004, \mnras, 351, 1277

\bibitem[{{Stellingwerf}(1978)}]{1978ApJ...224..953S}
{Stellingwerf}, R.~F. 1978, \apj, 224, 953

\bibitem[{{Takeda} {et~al.}(2019){Takeda}, {Han}, {Kang}, {Lee}, \&
  {Kim}}]{takeda19}
{Takeda}, Y., {Han}, I., {Kang}, D.-I., {Lee}, B.-C., \& {Kim}, K.-M. 2019,
  \mnras, 485, 1067

\bibitem[{{Tkachenko}(2015)}]{2015A&A...581A.129T}
{Tkachenko}, A. 2015, \aap, 581, A129

\bibitem[{{Tokovinin}(2008)}]{2008MNRAS.389..925T}
{Tokovinin}, A. 2008, \mnras, 389, 925

\bibitem[{{Torres} {et~al.}(1999){Torres}, {Lacy}, {Claret}, {Zakirov},
  {Arzumanyants}, {Bayramov}, {Hojaev}, {Stefanik}, {Latham}, \&
  {Sabby}}]{torres99}
{Torres}, G., {Lacy}, C. H.~S., {Claret}, A., {et~al.} 1999, \aj, 118, 1831

\bibitem[{{Warner} {et~al.}(2003){Warner}, {Kaye}, \&
  {Guzik}}]{2003ApJ...593.1049W}
{Warner}, P.~B., {Kaye}, A.~B., \& {Guzik}, J.~A. 2003, \apj, 593, 1049

\bibitem[{{Worek}(1996)}]{1996PASP..108..962W}
{Worek}, T.~F. 1996, \pasp, 108, 962

\bibitem[{{Worthey}(2015)}]{worthey2015}
{Worthey}, G. 2015, \aap, 580, L5

\end{thebibliography}

\begin{appendix}

\section{Additional figures and tables}
\label{Appendix_figures}

\begin{figure*}
\centering
\hspace{-.5cm}
\includegraphics[width=8.5cm]{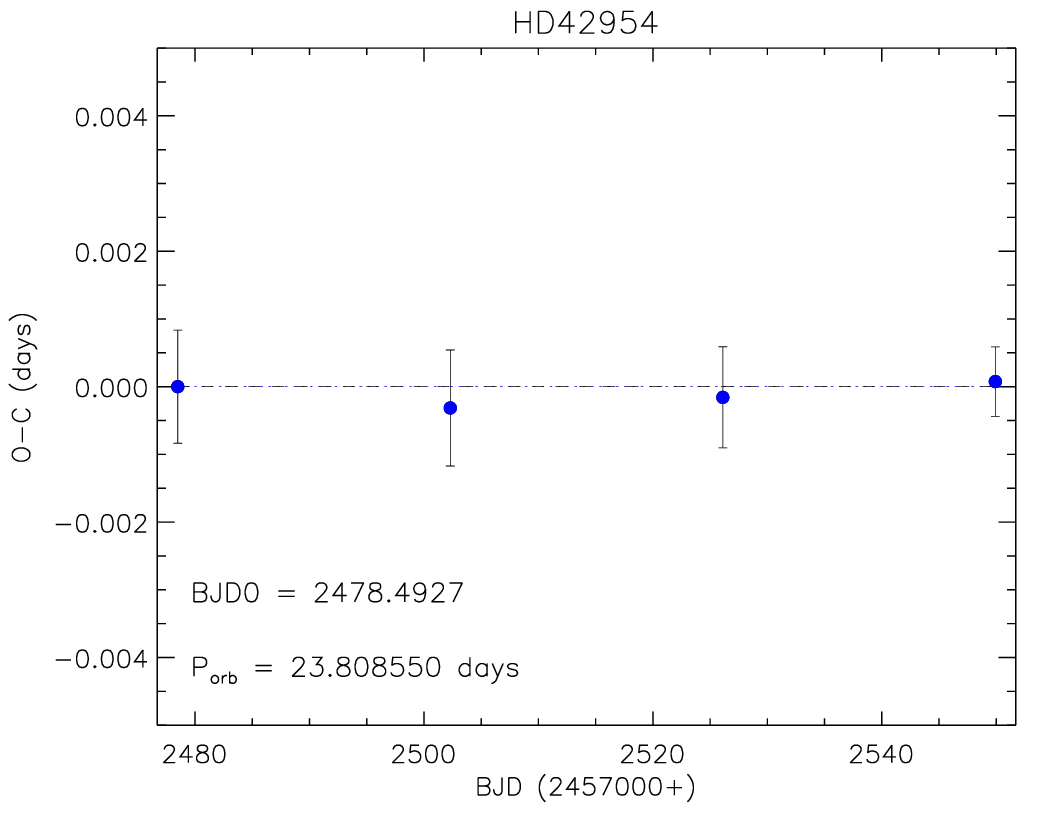}  
\includegraphics[width=8.5cm]{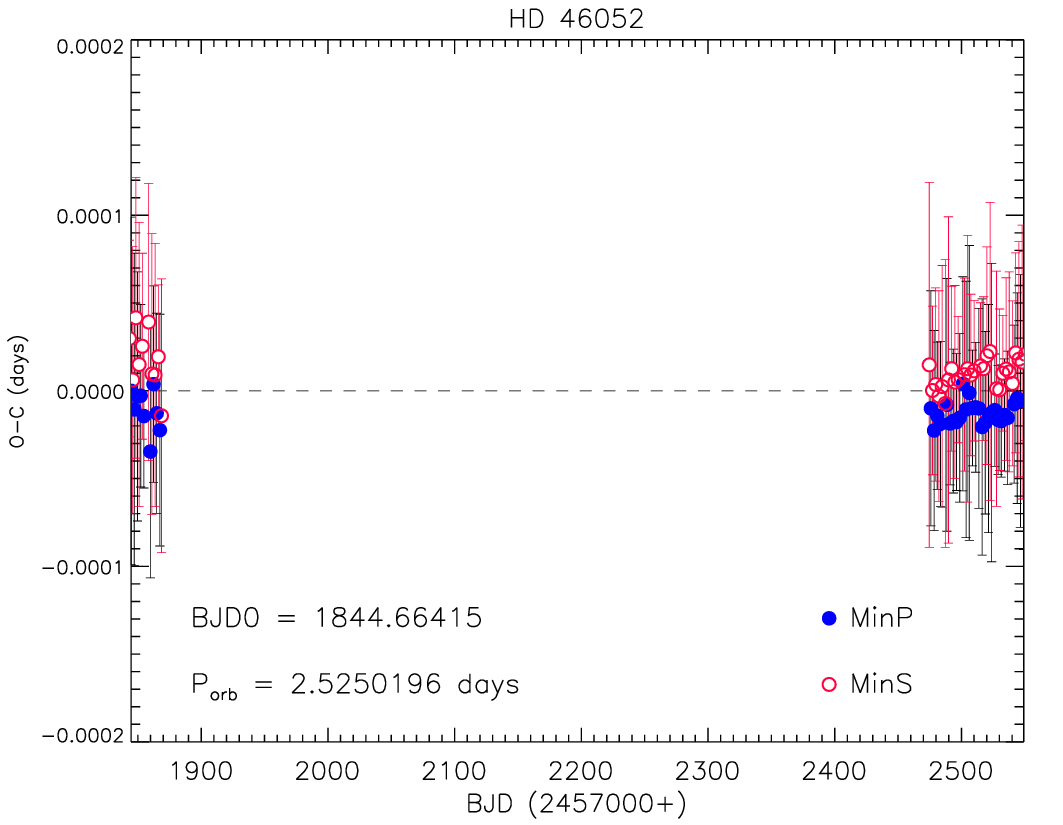}  
\includegraphics[width=8.5cm]{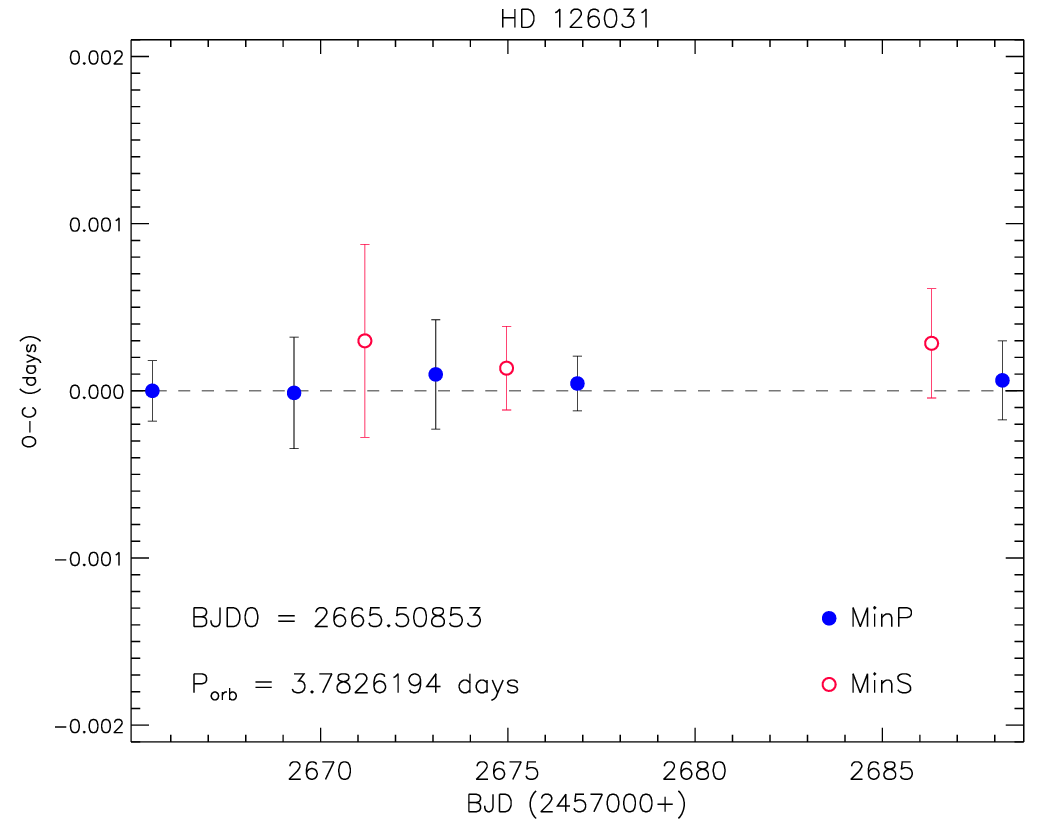}  
\includegraphics[width=8.5cm]{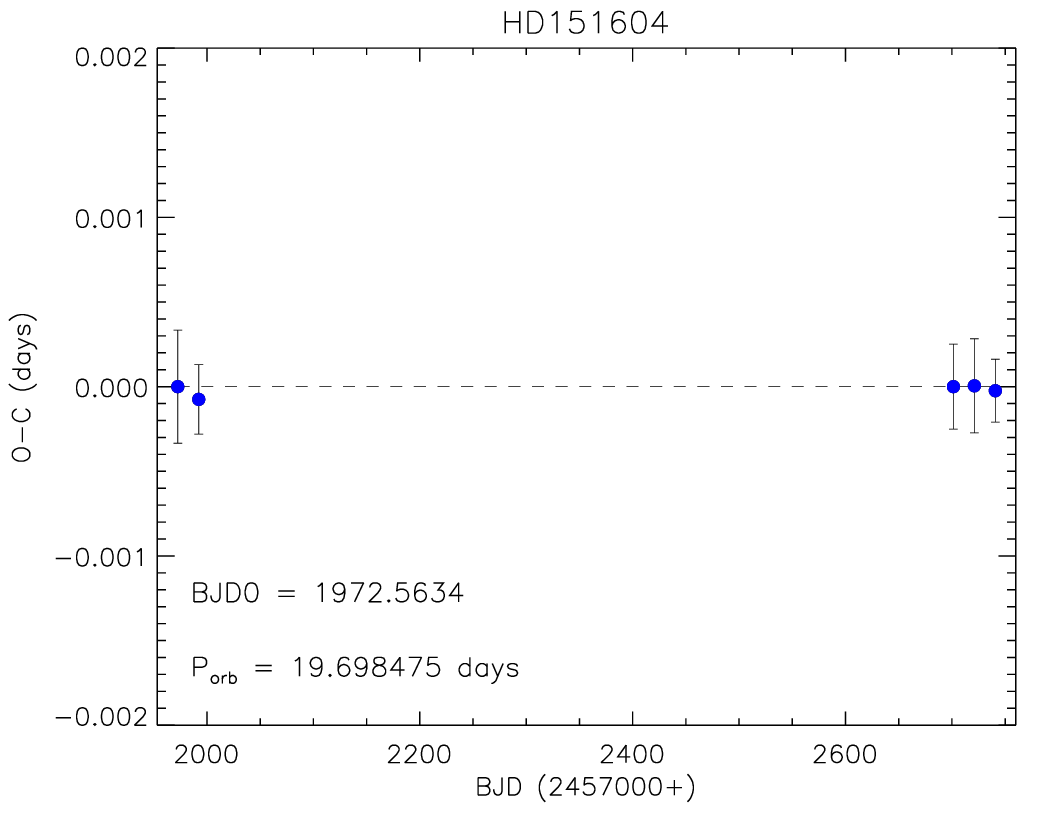}  
\includegraphics[width=8.5cm]{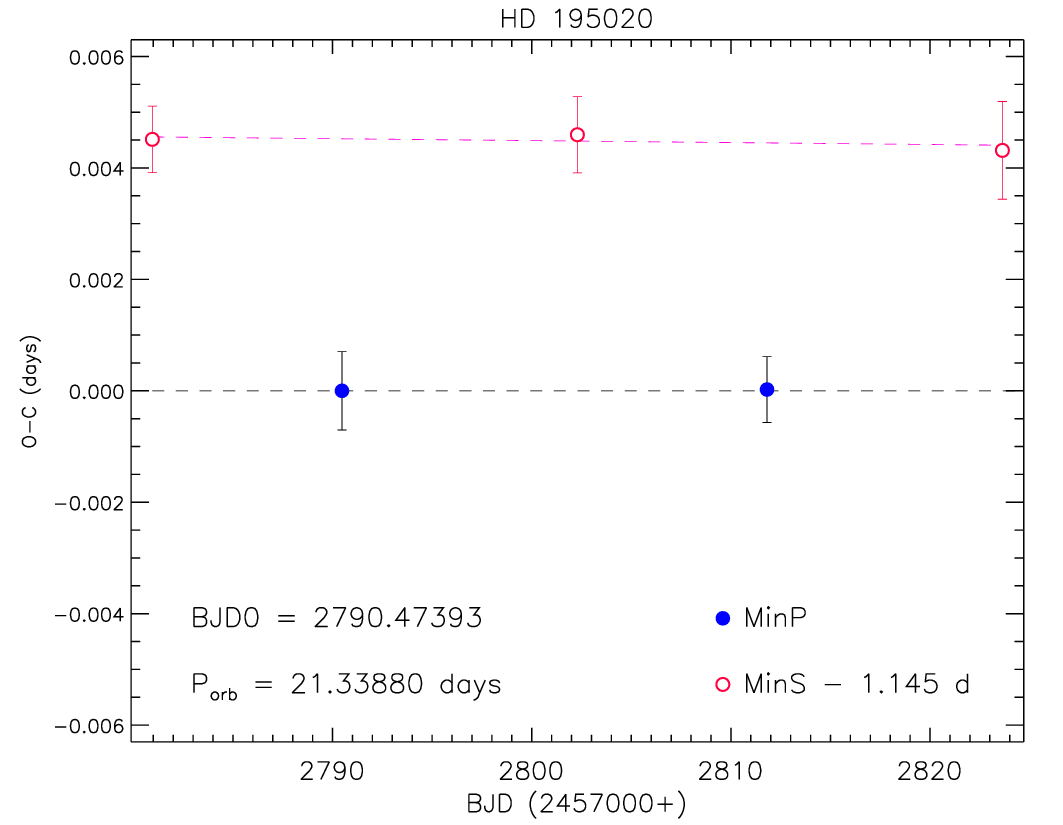}  
\includegraphics[width=8.5cm]{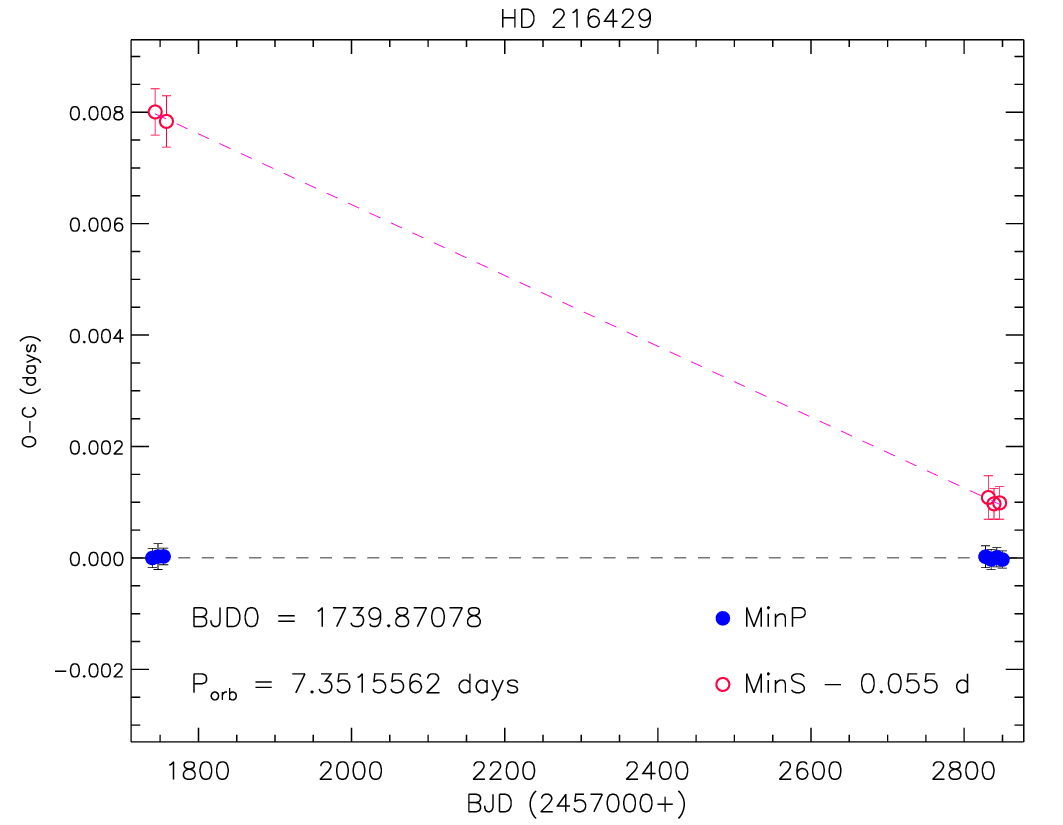}  
\vspace{0.cm}
\caption{Observed minus calculated times of minima for the six binaries. Primary and secondary minima (whenever observable) are represented with blue dots and red circles respectively. 
}
\label{Fig:o-c}
\end{figure*}

\begin{figure*}
\centering
	\includegraphics[width=16cm]{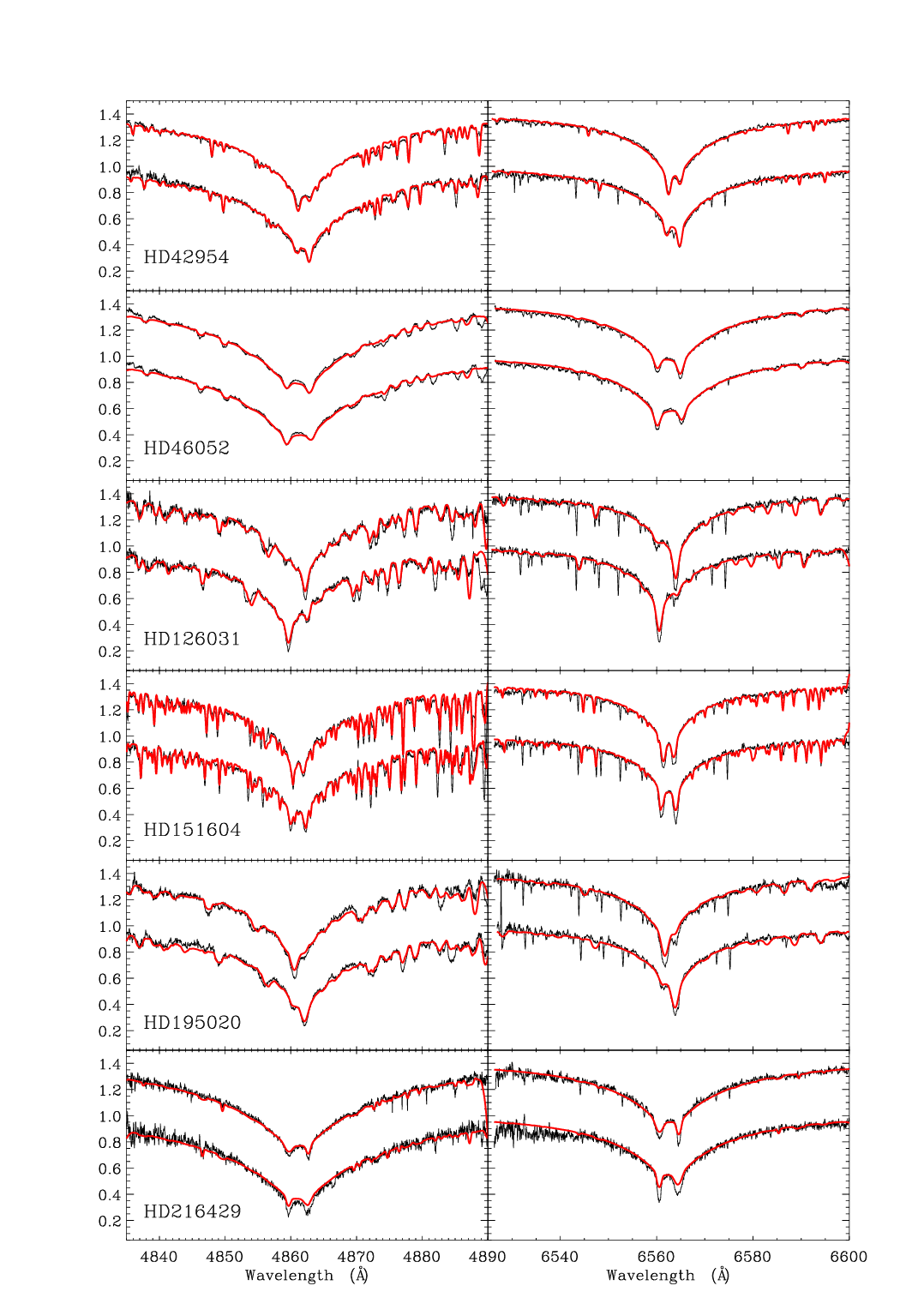}
    \caption{Results of the fitting procedure for the profiles of the Balmer lines, H$\beta$ (left panels) and H$\alpha$ (right panels), for the stars observed in this study. Two spectra corresponding to the quadrature orbital phases are displayed in each panel. The synthetic composite spectra are represented in red while the observed ones are shown in black.}
    \label{all_balmer}
\end{figure*}


\begin{table*}[thb]
\caption[]{Heliocentric Julian Date and measured radial velocities for SB2 stars of our sample.}
\label{rv_SB2}
\centering
\begin{tabular}{llrrllrr}
\hline
~~~~ID      &  ~~~~~JD        &     RV$_{\textrm{A}}$~~~~~ &     RV$_{\textrm{B}}$~~~~~  &~~~~   ID   &  ~~~~~JD        &     RV$_{\textrm{A}}$~~~~~ &     RV$_{\textrm{B}}$~~~~~  \\
            &  (2400000+)     & (km s$^{-1}$)~~ & (km s$^{-1}$)~~  &            &  (2400000+)     & (km s$^{-1}$)~~ & (km s$^{-1}$)~~  \\
\hline             
\hline
HD\,42954   & 56736.2954 & $-$16.85 $\pm$ 0.25   &      94.01 $\pm$ 1.20  &            & 58277.4627 & $-$52.41 $\pm$ 0.38   &     24.80 $\pm$ 0.42 \\
            & 56747.2782 &    64.27 $\pm$ 0.40   &    $-$4.06 $\pm$ 1.26  &            & 58290.4439 &  $-$5.65 $\pm$ 0.41   &  $-$22.13 $\pm$ 0.41 \\  
            & 57004.6426 &    97.78 $\pm$ 0.29   &   $-$44.86 $\pm$ 1.13  &            & 58298.4131 & $-$58.42 $\pm$ 0.42  &     33.56 $\pm$ 0.45 \\
            & 57014.4925 &    33.30 $\pm$ 0.16   &      33.30 $\pm$ 0.16  &            & 58299.4475 & $-$65.13 $\pm$ 0.39  &     37.38 $\pm$ 0.42 \\
            & 57043.4847 &  $-$0.50 $\pm$ 0.47   &      73.27 $\pm$ 1.34  &            & 58305.4251 &    55.31 $\pm$ 0.30  &  $-$82.42 $\pm$ 0.30 \\
            & 57332.6198 & $-$26.01 $\pm$ 0.50   &     103.23 $\pm$ 1.69  &            & 58306.4416 &    34.90 $\pm$ 0.30  &  $-$66.53 $\pm$ 0.34 \\
            & 59211.4345 &  $-$7.46 $\pm$ 1.41   &      87.73 $\pm$ 3.62  &            & 58311.3967 & $-$14.51 $\pm$ 0.13  &  $-$14.51 $\pm$ 0.13 \\
            & 59212.4532 & $-$15.87 $\pm$ 1.28   &      96.08 $\pm$ 3.41  &            & 58312.4006 & $-$23.20 $\pm$ 0.27  &   $-$3.85 $\pm$ 0.31 \\
            & 59222.4716 &    71.91 $\pm$ 1.00   &   $-$10.98 $\pm$ 3.15  &            & 58319.3754 & $-$65.70 $\pm$ 0.30  &     38.45 $\pm$ 0.34 \\
            & 59223.3465 &    65.83 $\pm$ 0.98   &    $-$2.24 $\pm$ 2.32  &            & 59299.6584 & $-$39.14 $\pm$ 0.99  &     12.51 $\pm$ 1.03 \\
            & 59242.2640 &    99.83 $\pm$ 2.43   &   $-$42.80 $\pm$ 5.98  &            & 59300.6572 & $-$46.29 $\pm$ 0.51  &     17.27 $\pm$ 0.54 \\
            & 59250.2964 &    46.64 $\pm$ 0.91   &      22.94 $\pm$ 2.02  &            & 59313.6255 &     4.22 $\pm$ 0.38  &  $-$33.19 $\pm$ 0.42 \\
            & 59251.3107 &    38.11 $\pm$ 0.45   &      24.10 $\pm$ 5.45  &            & 59314.6168 &  $-$7.16 $\pm$ 1.54  &  $-$19.12 $\pm$ 1.20 \\
            & 59263.2457 & $-$14.78 $\pm$ 0.45   &      93.88 $\pm$ 1.52  &            & 59315.6081 & $-$14.19 $\pm$ 0.45  &  $-$14.19 $\pm$ 0.45 \\
            & 59263.5132 &  $-$4.90 $\pm$ 0.53   &      81.81 $\pm$ 1.55  &            & 59353.5277 &  $-$0.34 $\pm$ 0.86  &  $-$27.30 $\pm$ 0.92 \\
            & 59264.2828 &    40.62 $\pm$ 0.73   &      17.73 $\pm$ 3.49  &            & 59419.4056 & $-$46.87 $\pm$ 1.27  &     23.37 $\pm$ 1.32 \\
            & 59264.5061 &    56.02 $\pm$ 2.31   &       8.47 $\pm$ 5.00  &            & 59421.3652 & $-$61.22 $\pm$ 0.36  &     32.17 $\pm$ 0.40 \\
            & 59271.2567 &    64.32 $\pm$ 1.43   &       0.57 $\pm$ 3.14  &            & 59424.3754 & $-$60.95 $\pm$ 1.66  &     31.97 $\pm$ 1.77 \\
            & 59272.3073 &    58.05 $\pm$ 1.35   &       8.08 $\pm$ 3.16  &            & 59425.3644 & $-$14.95 $\pm$ 1.51  &  $-$14.95 $\pm$ 1.51 \\
            & 59300.3222 &    32.78 $\pm$ 3.27   &      32.78 $\pm$ 3.27  &            & 59426.3738 &    64.82 $\pm$ 1.65  &  $-$93.71 $\pm$ 1.73 \\
            & 59314.2900 &    98.88 $\pm$ 1.09   &   $-$42.01 $\pm$ 2.85  &            &            &                      &                     \\ 
            & 59315.3117 &    91.37 $\pm$ 2.84   &   $-$33.95 $\pm$ 6.62  & HD\,195020 & 56875.4805 &  $-$50.89 $\pm$ 1.11 &    53.56 $\pm$ 2.93  \\
            &            &                       &                        &            & 58291.5841 &     47.68 $\pm$ 1.18 & $-$76.82 $\pm$ 3.43  \\
HD\,46052   & 56737.2874 &  $-$77.18 $\pm$ 0.76  &      61.95 $\pm$ 1.13  &            & 58299.5564 &  $-$27.38 $\pm$ 1.13 &    25.54 $\pm$ 4.43  \\
            & 57014.5086 & $-$119.99 $\pm$ 1.41  &     107.21 $\pm$ 1.75  &            & 58306.5219 &  $-$40.91 $\pm$ 0.54 &    39.98 $\pm$ 1.72  \\
            & 57043.5060 &     94.53 $\pm$ 0.78  &  $-$121.76 $\pm$ 1.13  &            & 58311.5487 &     46.49 $\pm$ 1.26 & $-$77.40 $\pm$ 1.32  \\
            & 59211.5171 & $-$122.37 $\pm$ 4.48  &     113.04 $\pm$ 5.59  &            & 58330.5347 &      6.74 $\pm$ 1.74 & $-$21.96 $\pm$ 5.24  \\
            & 59212.5137 &     98.13 $\pm$ 6.73  &  $-$122.96 $\pm$ 8.13  &            & 58367.4846 &  $-$50.53 $\pm$ 1.76 &    56.20 $\pm$ 4.22  \\
            & 59222.4468 &     68.54 $\pm$ 4.02  &   $-$94.08 $\pm$ 5.03  &            & 59419.5621 &     32.71 $\pm$ 1.38 & $-$47.71 $\pm$ 2.32  \\
            & 59223.3949 &  $-$10.51 $\pm$ 0.84  &   $-$10.51 $\pm$ 0.84  &            & 59420.5685 &     43.99 $\pm$ 2.70 & $-$70.68 $\pm$ 5.31  \\
            & 59242.4479 &     28.04 $\pm$ 2.38  &   $-$40.10 $\pm$ 4.27  &            & 59423.5171 &     41.33 $\pm$ 6.88 & $-$65.04 $\pm$ 6.57  \\
            & 59250.3784 &     98.12 $\pm$ 0.94  &  $-$122.46 $\pm$ 1.26  &            & 59425.5776 &     20.56 $\pm$ 6.77 & $-$41.49 $\pm$ 6.64  \\
            & 59251.4048 &  $-$70.93 $\pm$ 1.89  &      56.50 $\pm$ 2.47  &            & 59449.4996 &   $-$7.91 $\pm$ 0.23 &  $-$7.91 $\pm$ 0.23  \\
            & 59263.3612 &     95.67 $\pm$ 2.99  &  $-$121.30 $\pm$ 3.54  &            & 59454.3823 &  $-$45.09 $\pm$ 0.61 &    48.01 $\pm$ 1.78  \\
            & 59264.3716 & $-$125.38 $\pm$ 1.00  &     113.96 $\pm$ 1.39  &            & 59497.3700 &  $-$52.45 $\pm$ 2.66 &    44.01 $\pm$ 4.31  \\
            & 59272.3251 &  $-$91.58 $\pm$ 1.98  &      82.67 $\pm$ 2.44  &            & 59794.4428 &  $-$35.44 $\pm$ 1.34 &    38.99 $\pm$ 2.47  \\
            & 59299.3989 &  $-$76.17 $\pm$ 5.37  &      62.63 $\pm$ 6.39  &            & 59795.4282 &  $-$42.84 $\pm$ 1.58 &    45.86 $\pm$ 2.56  \\
            & 59300.3456 &  $-$32.78 $\pm$ 2.55  &      20.71 $\pm$ 3.17  &            & 59800.3605 &  $-$37.32 $\pm$ 2.42 &    40.12 $\pm$ 3.71  \\
            & 59314.3066 &   $-$9.10 $\pm$ 0.91  &    $-$9.10 $\pm$ 0.91  &            & 59801.3768 &  $-$30.79 $\pm$ 1.91 &     4.40 $\pm$ 2.06  \\ 
            & 59315.3230 &  $-$79.43 $\pm$ 5.09  &      65.18 $\pm$ 6.38  &            & 59838.3832 &  $-$43.76 $\pm$ 1.82 &    47.81 $\pm$ 2.93  \\
            &            &                       &                        &            &            &                      &                      \\
HD\,126031  & 56802.3688 &  $-$71.02 $\pm$ 0.80  &     27.31 $\pm$  3.44  & HD\,216429 & 56875.5465 &    18.88 $\pm$  5.18 &  $-$46.70 $\pm$ 1.20 \\
            & 59420.3018 & $-$102.48 $\pm$ 0.61  &     69.85 $\pm$  2.43  &            & 59419.6062 &    21.70 $\pm$  7.65 &  $-$18.07 $\pm$ 3.34 \\
            & 59421.3175 &  $-$54.92 $\pm$ 1.76  &      9.28 $\pm$  5.06  &            & 59423.5987 &    14.43 $\pm$  9.74 &  $-$20.61 $\pm$ 6.68 \\        
            & 59422.3483 &     52.25 $\pm$ 0.80  & $-$133.06 $\pm$  4.23  &            & 59435.6119 & $-$48.68 $\pm$  9.95 &     38.49 $\pm$ 1.69 \\
            & 59423.3436 &  $-$20.18 $\pm$ 2.00  &  $-$33.18 $\pm$  2.04  &            & 59436.6059 & $-$95.84 $\pm$  7.94 &     77.80 $\pm$ 3.92 \\
            & 59424.3049 & $-$110.63 $\pm$ 4.50  &     78.33 $\pm$ 13.76  &            & 59442.5765 & $-$31.90 $\pm$  8.83 &     18.70 $\pm$ 4.57 \\
            & 59426.3220 &     53.37 $\pm$ 2.42  & $-$131.81 $\pm$ 11.11  &            & 59444.5764 &$-$104.80 $\pm$  8.65 &     83.64 $\pm$ 4.90 \\
            & 59748.3271 &     21.13 $\pm$ 0.96  &  $-$93.28 $\pm$  3.72  &            & 59445.5508 & $-$33.37 $\pm$  8.86 &      1.78 $\pm$ 5.14 \\
            & 59755.3710 &     53.68 $\pm$ 1.08  & $-$135.48 $\pm$  4.65  &            & 59450.6031 & $-$69.56 $\pm$ 10.22 &     49.17 $\pm$ 4.68 \\
            & 59757.3624 & $-$107.61 $\pm$ 0.98  &     76.91 $\pm$  4.46  &            & 59470.4906 &    28.93 $\pm$  6.78 &  $-$48.70 $\pm$ 6.04 \\
            & 59768.3219 & $-$102.33 $\pm$ 0.90  &     70.54 $\pm$  3.23  &            & 59471.4998 & $-$40.08 $\pm$  7.20 &   $-$4.12 $\pm$ 4.79 \\
            & 59776.3817 & $-$103.92 $\pm$ 0.92  &     70.84 $\pm$  3.63  &            & 59472.5733 & $-$65.83 $\pm$  8.60 &     45.00 $\pm$ 6.01 \\
            & 59777.3589 &      7.49 $\pm$ 0.92  &  $-$76.36 $\pm$  3.69  &            & 59497.4366 &    31.16 $\pm$ 12.83 &  $-$67.53 $\pm$ 6.15 \\
            & 59786.3513 &   $-$2.14 $\pm$ 0.91  &  $-$62.94 $\pm$  3.17  &            & 59498.4837 &    77.41 $\pm$ 12.54 & $-$104.31 $\pm$ 6.08 \\
            & 59787.3501 & $-$108.31 $\pm$ 0.89  &     75.58 $\pm$  4.28  &            & 59794.5621 &    18.91 $\pm$  6.14 &  $-$20.69 $\pm$ 1.12 \\
            & 59800.2854 &     32.44 $\pm$ 0.94  & $-$110.96 $\pm$  3.79  &            & 59795.5734 & $-$34.60 $\pm$  7.80 &     24.89 $\pm$ 1.69 \\
            &            &                       &                        &            & 59818.5032 & $-$83.80 $\pm$  5.77 &     63.64 $\pm$ 1.57 \\
HD\,151604  & 56820.4401 & $-$57.42 $\pm$ 0.36   &     27.61 $\pm$ 0.38   &            & 59819.4867 & $-$102.54$\pm$  6.67 &     84.37 $\pm$ 1.66 \\
            & 58269.4793 &     5.83 $\pm$ 0.45   &  $-$34.32 $\pm$ 0.50   &            & 59820.5178 & $-$34.65 $\pm$ 3.453 &  $-$2.03  $\pm$ 2.43 \\
            & 58270.4782 &  $-$3.59 $\pm$ 0.42   &  $-$23.98 $\pm$ 0.43   &            &            &                      &                      \\
\hline
\\
\end{tabular}
\end{table*}

\begin{table*}
\caption{Chemical abundances inferred for the stars of our sample. All the values are referred to the solar ones \citep{2010Ap&SS.328..179G}. Typical uncertainties are $\pm$0.2\,dex.}

\label{tab:abun}
\centering
\begin{tabular}{lrrrrrrrrrrrr}
\hline
\hline
 El & \multicolumn{2}{r}{HD\,42954} &\multicolumn{2}{r}{HD\,46052} &\multicolumn{2}{r}{HD\,126031} &\multicolumn{2}{r}{HD\,151604} &\multicolumn{2}{r}{HD\,195020} &\multicolumn{2}{r}{HD\,216429} \\
   & A & B & A & B & A & B & A & B & A & B & A & B \\
\hline
C  & $-$1.0 & $-$1.0 & $-$0.3 & $-$0.5 & $-$1.0 &    0.0 & $-$1.2 & $-$1.2  & $-$1.0 & $-$0.5 & $-$0.3 & $-$0.7  \\
O  & $-$0.8 & $-$1.3 &    0.4 &    0.4 &    0.0 &    0.0 & $-$0.3 & $-$0.3  &    0.0 &    0.0 &    1.2 &    1.4  \\
Na & $-$0.7 & $-$1.3 &    0.5 &    0.5 &    0.7 &    0.5 &    0.5 &    0.6  & $-$0.3 &    0.4 &    0.0 &    0.0  \\
Mg & $-$1.0 & $-$2.0 &    0.5 &    0.5 &    0.3 &    0.0 &    0.4 &    0.4  & $-$0.2 &    0.0 & $-$0.7 & $-$1.0  \\
Si & $-$0.1 & $-$0.8 &    0.3 &    0.3 &    0.5 &    0.5 &    0.3 &    0.3  & $-$0.5 &    0.5 & $-$0.7 & $-$1.0  \\
Ca & $-$0.8 & $-$0.8 & $-$0.2 & $-$0.2 & $-$0.6 & $-$0.1 & $-$0.6 & $-$0.8  & $-$0.1 & $-$0.5 & $-$0.5 &    0.0  \\
Sc & $-$1.6 & $-$1.0 & $-$0.6 & $-$0.6 & $-$1.1 & $-$0.1 & $-$1.8 & $-$1.5  & $-$1.0 & $-$0.7 & $-$0.1 & $-$0.1  \\
Ti & $-$0.8 & $-$1.6 &    0.8 &    0.8 &    0.8 &    0.5 &    0.5 &    0.5  & $-$0.1 &    0.8 &    0.0 &    0.0  \\
Cr &    0.0 & $-$1.0 &    0.5 &    0.5 &    0.5 &    0.0 &    0.4 &    0.4  & $-$0.2 &    0.4 & $-$0.1 & $-$0.1  \\
Mn &    0.0 &    0.0 &    0.0 &    0.5 &    0.5 &    0.0 &    0.3 &    0.0  &    0.0 &    0.0 &    0.0 &    0.0  \\
Fe & $-$0.3 & $-$0.2 &    0.5 &    0.7 &    0.5 &    0.3 &    0.3 &    0.3  &    0.0 &    0.4 &    0.0 &    0.0  \\
Ni &    0.2 &    0.2 &    0.2 &    0.4 &    0.8 &    0.0 &    0.6 &    0.6  &    0.0 &    0.4 &    0.0 &    0.0  \\
Zn &    0.1 & $-$0.1 &    1.0 &    1.0 &    0.6 &    0.0 &    0.6 &    0.6  & $-$0.6 &    0.2 &    0.0 &    0.0  \\
Sr &    0.0 & $-$1.6 &    2.0 &    2.0 &    1.6 &    0.0 &    0.6 &    0.6  &    0.0 &    1.1 &    0.0 &    0.6  \\
Y  &    0.2 &    0.0 &    0.1 &    0.1 &    1.3 &    0.0 &    0.6 &    0.0  & $-$0.7 &    0.3 &    0.0 &    0.0  \\
Zr &    0.2 &    0.2 &    1.4 &    1.0 &    2.0 &    0.0 &    0.9 &    0.9  &    0.0 &    0.0 &    0.0 &    0.6  \\
Ba &    0.4 &    0.1 &    2.4 &    2.4 &    1.9 &    0.9 &    1.9 &    1.9  &    0.3 &    1.6 &    0.4 &    0.9  \\
\hline

\end{tabular}
\end{table*}

\begin{figure*} 
\begin{center}
\hspace{-.5cm}	
\includegraphics[width=\columnwidth]{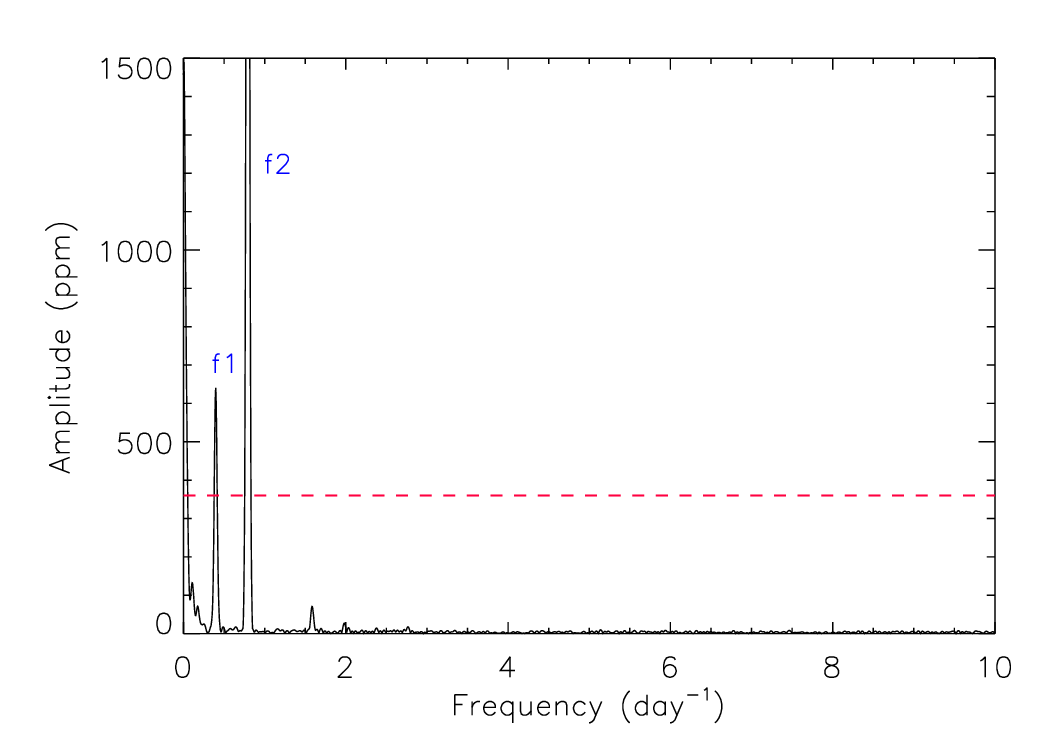}
\includegraphics[width=\columnwidth]{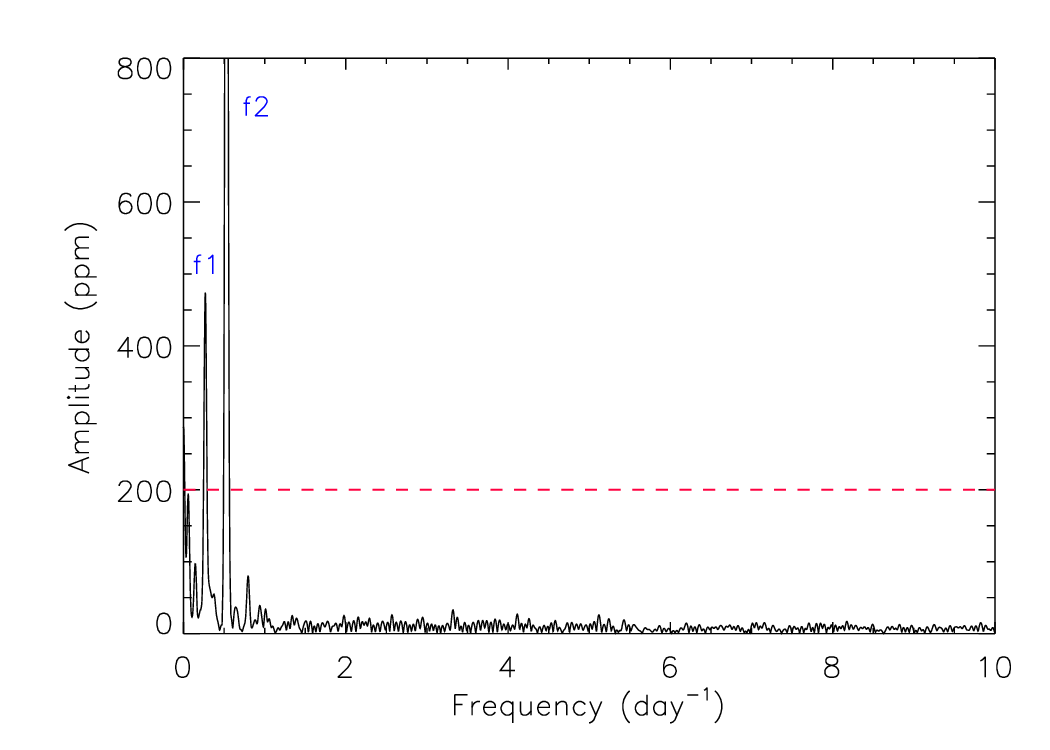}
\includegraphics[width=\columnwidth]{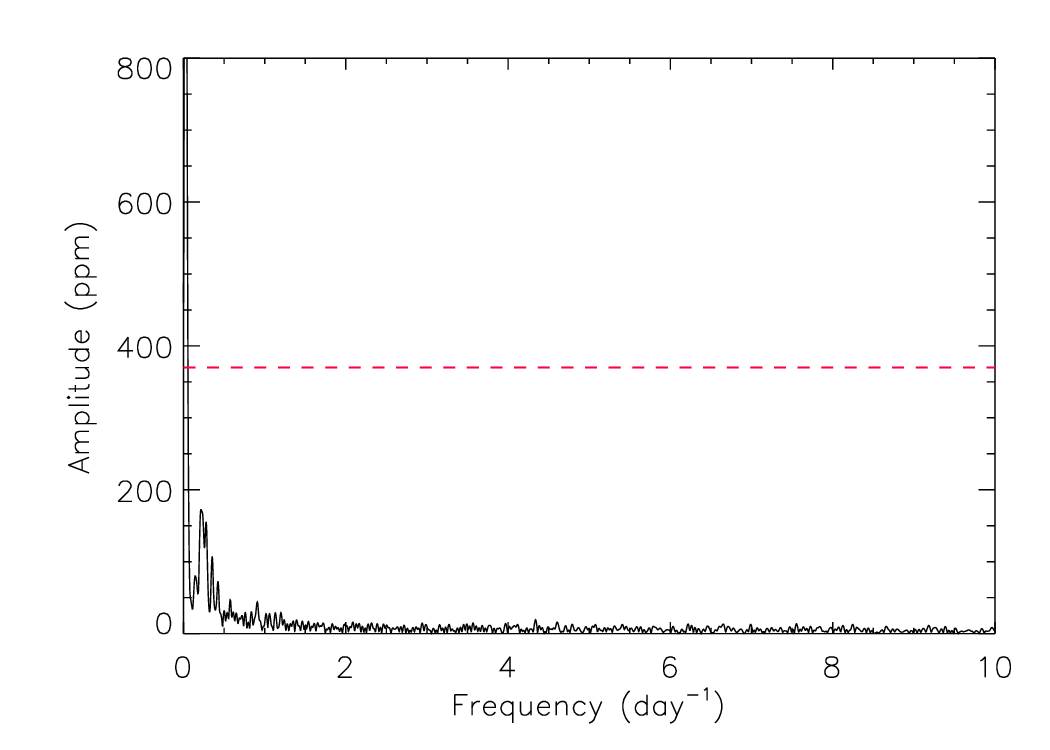}
\vspace{0cm}
\caption{Cleaned periodogram of the \tess\ light curve of HD\,46052 (left upper panel) taken in Sector 43 (September--October 2021), HD\,126031 (right upper panel) taken in Sector 50 (March--April 2022), and HD\,195020 (center bottom panel) taken in Sector 55 (August 2022), where the eclipses have been removed. The only two peaks higher than 99\% confidence level (red dashed line) correspond to the orbital period and its half and are related to proximity effects (reflection and ellipsoidal shape of the components).}
\label{fig:freq_HD46052_HD126031}
\end{center}
\end{figure*}

\end{appendix}

\end{document}